\documentclass[review]{elsarticle}

\usepackage{lineno,hyperref}
\modulolinenumbers[5]

\journal{Neural Networks}

\biboptions{authoryear}

\usepackage{epsfig}
\graphicspath{{figs/}}

\usepackage{amsmath}
\begin{document}

\begin{frontmatter}

\title{Online regularization of complex-valued neural networks for structure optimization in wireless-communication channel prediction
}

\author[UTAffiliation,WUSTLAffiliation]{Tianben Ding}
\ead{tding@wustl.edu}

\author[UTAffiliation]{Akira Hirose\corref{mycorrespondingauthor}}
\cortext[mycorrespondingauthor]{Correspondence author.}
\ead{ahirose@ee.t.u-tokyo.ac.jp}

\address[UTAffiliation]{Department of Electrical Engineering and Information Systems, The University of Tokyo, Tokyo, Japan}
\address[WUSTLAffiliation]{Department of Electrical and Systems Engineering, Washington University in St. Louis, MO, United States}




\begin{abstract}
This paper proposes online-learning complex-valued neural networks (CVNNs) to predict future channel states in fast-fading multipath mobile communications. CVNN is suitable for dealing with a fading communication channel as a single complex-valued entity. This framework makes it possible to realize accurate channel prediction by utilizing its high generalization ability in the complex domain. However, actual communication environments are marked by rapid and irregular changes, thus causing fluctuation of communication channel states. Hence, an empirically selected stationary network gives only limited prediction accuracy. In this paper, we introduce regularization in the update of the CVNN weights to develop online dynamics that can self-optimize its effective network size by responding to such channel-state changes. It realizes online adaptive, highly accurate and robust channel prediction with dynamical adjustment of the network size. We demonstrate its online adaptability in simulations and real wireless-propagation experiments.
\end{abstract}


\end{frontmatter}

\linenumbers

\section{Introduction}
Performance of mobile communications always suffers from signal degradation, namely fading, due to path loss, shadowing, interference and channel state changes caused by movement of users \citep{Cho2010}. In principle, fading can be mitigated by pre-equalization such as zero-forcing \citep{Ho2017} or minimum-mean-square-error (MMSE) equalization \citep{Eraslan2013}. Transmission power control is another countermeasure against the fading phenomenon \citep{Ren2018}. These methods rely on accurate estimation of channel state information at communication ends. However, in practical mobile communications, the channel state, or simply channel, changes rapidly and irregularly due to time-varying multipath environments caused by movement of mobile users and their surroundings. The time fluctuation outdates the estimated channel and degrades the communication quality significantly. Channel prediction is an effective way to overcome this problem by forecasting channel changes over time based on preceding information. An accurate channel prediction is required for communication quality and adaptive transmission in the next-generation communications \citep{Duel-Hallen2007,Bui2017}. 

Several articles exist on the channel prediction in mobile communications including, for example, methods based on linear \citep{Maehara2003,Bui2013} and autoregressive (AR) model extrapolation \citep{Eyceoz1998,Arredondo2002,Duel-Hallen2006,Sharma2007}. Although the low computational complexity in these methods is suitable for real-time operation in mobile communications, such simple linear or AR-model-based methods provide limited performance on predicting rapid channel changes \citep{Ding2014b}. Neural-network-based channel prediction methods have also attracted attention due to the recent successful development of artificial neural networks in various engineering fields. The generalization ability of neural networks provides flexible representation of complicated channel-state changes and high prediction capability. For instance, prediction methods based on an echo-state-network (ESN) \citep{Zhao2017} and an extreme-learning-machine (ELM) \citep{Sui2018} as well as real-valued recurrent-neural-network (RNN) \citep{Liu2006,Potter2010} have been reported, and their prediction performance has been evaluated in some simulated communication situations. Luo \textit{et al.} recently combined a convolutional neural network and a long short-term memory (LSTM) network for learning and predicting channel states under specific communication situations \citep{Luo2018}. To realize a high-precision prediction in practical mobile communications, the authors also proposed a method based on a multiple-layer complex-valued neural network (ML-CVNN) by focusing rotary motion of the channel state in the complex plane. This method led to superior channel prediction performance in some simulated and practical communication scenarios \citep{Ding2014b}.

Generally, in applications of neural networks, size of networks is critical to the application performance because it affects the generalization characteristics and calculation costs \citep{Hirose2012,Ramachandram2017}. For example, a too small network is not enough to represent the complexity of targets, resulting low convergence properties. On the other hand, a too large network requires expensive calculation costs, and most importantly, it causes overfitting. Despite its importance, the structure of the network is typically defined based on a rule of thumb by users. One may start with an arbitrary structure and evaluate its learning performance using a large amount of training data by increasing or decreasing the number of neurons and network connections until the best structure is found. This is also true in the state-of-the-art neural-network-based methods in channel prediction. For example, in our previous prediction method, we empirically set the structure of the CVNN (number of fully connected input terminals and neurons in a hidden layer) based on its prediction accuracy in a series of simulated communication situations. Although the structure shows a high prediction performance on some simulated and experimentally observed fading channels \citep{Ding2014b}, this manual pre-tuning of the network parameters is time consuming and inefficient. Moreover, mobile communications in the real world is forced to work in more diverse communication environments and experiences more rapid and various fluctuations than those in simplified simulations. As a result, an \textit{a priori} tuned structure under a situation is no longer optimal for other practical communication environments. The most suitable neural-network structure in a channel prediction method should also be dynamical according to the changes of communication environments. This motivates us to develop a neural-network scheme optimizing network structures dynamically and adaptively for the channel prediction. 

 In this paper, in order to realize a dynamically optimized network structure to suit best to the fading channel at each moment, we propose a new ML-CVNN-based channel prediction method by introducing regularization. We work with a large-size network platform and then let it automatically find, or self-adjust to, a suitable structure within the platform that uses only a limited portion of the network in order to achieve a good generalization. The self-adjustment is performed by imposing sparse constraints \citep{Tibshirani1996,Elad2010} to the connection weight updates. The sparse constraints suppress the redundant connection weights to be zero, and equivalently construct a smaller scaled network using only the remaining non-zero connections \citep{Ding2014a}. Although introduction of sparsity in neural network is a relatively standard strategy for reduction of overfitting problem \citep{Aghasi2017} and/or computational requirement upon practical implementation \citep{Scardapane2017,Koneru2019}, it has not been carefully discussed in the literature of complex-valued neural networks and considered in the channel prediction. Here, we introduce $L _1$-norm (Lasso) and $L _{2,1}$-norm (group-Lasso) penalty into ML-CVNN updates as the constraints and validate performance of weight-level and neuron-level sparsity in the channel prediction context. In order to follow the time fluctuation in the channel state, we develop an online training-and-prediction framework. We update the network by using a set of the most recent channels immediately before the prediction with a small number of learning iterations. We keep the updated network structure temporarily for the next training-and-predicting time frame. In this way, the non-zero connection distribution changes from time to time in the structure so that it keeps the most suitable size of the network for each prediction situation. 

In each training phase, we use a backpropagation of teacher signal (BPTS) \citep{Hirose1996}, rather than the standard error-backpropagation. The BPTS-based update method is simpler and has a lower computational cost, which is preferred for mobile communications. We demonstrate that the new channel prediction methods with the online adaptive CVNN structures present highly accurate predictions under fluctuating communication environments in a series of simulations and experiments. Further, we closely observe and discuss the effects of the dynamically changing structures on the bit-error rate performance.
 
 The major contributions of our study can be summarized as follows:
 \begin{enumerate}
    \item Proposal of complex-valued update schemes that self-adjust network structures to provide suitable network size by responding complicated channel states;
    \item Design of new channel prediction methods based on dynamic ML-CVNNs with the proposed network structures and the BPTS for an adaptive prediction;
    \item Verification of the fact that the proposed fast fading prediction has a performance superior to other approaches on simulated and experimentally observed channel states.
\end{enumerate}

This paper is organized as follows: Section~\ref{sec:channelModel} briefly introduces the channel model theory and path separation in the frequency domain. After reviewing the conventional CVNN-based channel prediction in Section~\ref{sec:conventionalCVNN}, we propose a novel prediction method based on a ML-CVNN with the dynamically changing structure in Section~\ref{sec:dynamicCVNN}. Then, Sections~\ref{sec:simulation} and \ref{sec:experiment} present its performance in simulations and experiments, respectively. Finally, Section~\ref{sec:conclusion} provides the conclusion.

\section{Channel Model and Multipath Separation in Frequency Domain}
\label{sec:channelModel}

\begin{figure}[!t]
\centering
\includegraphics[width=3.33in]{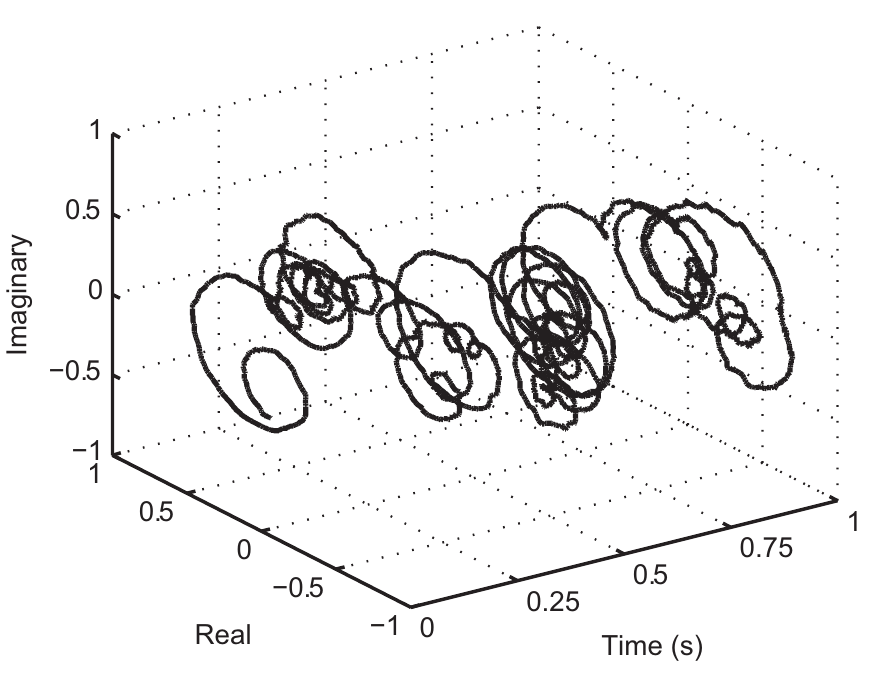}
\caption{An example of time-varying fading channel states in the complex domain measured in an actual mobile communication.}
\label{fig:channelChangeExp}
\end{figure}

\begin{figure}[!t]
\centering
\includegraphics[width=3.33in]{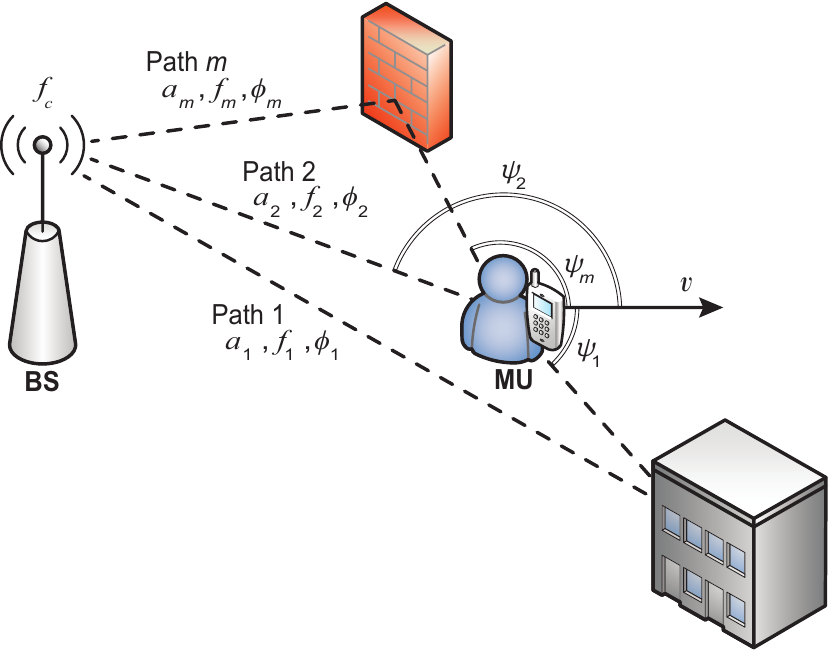}
\caption{Jakes multipath model. Communication channel between a base station (BS) and a mobile user (MU) is distorted by interference of multipath and movement of communication ends and/or scatterers.}
\label{fig:multipathModel}
\end{figure}

Channel states of communications are distorted mainly by multipath interference caused by scattering in the communication environment. In addition, movement of mobile users and/or scatterers causes rapid and irregular channel changes in time. Fig.~\ref{fig:channelChangeExp} shows an example of fading channel states in actual mobile communications. The curves demonstrate irregularity and nonlinearity of channel changes in the complex domain, and express difficulty of channel prediction because of its irregularly rotation-like changes. Generally, a signal received at a communication end $y (t)$ at time $t$ is modeled with time-varying channel $c (t)$ as
\begin{equation}\label{eq:receivedSig}
y ( t ) = c (t) s (t) + n (t) 
\end{equation}
where $s (t)$ and $n (t)$ are transmitted signal and additive white Gaussian noise (AWGN), respectively. According to the Jakes model \citep{Jakes1994,Cho2010}, fading channel $c ( t )$ as a function of time $t$ can be modeled as a summation of individual $M$ complex signal paths $c _m(t)$ at a receiver and expressed as
\begin{equation}\label{eq:channelState}
c ( t ) = \sum _{m=1}^{M}c_{m}(t)= \sum _{m = 1} ^{M} a _{m} e ^{j ( 2 \pi f _{m} t + \phi _{m} )}
\end{equation}
where $a _{m}$, $f _{m}$, and $\phi _{m}$ are amplitude, Doppler frequency, and phase shift of each single path $m$, and $M$ is the total path number. The Doppler frequency due to movement of a mobile user is given by 
\begin{equation}\label{eq:DopplerFre}
f _m = \frac{f_c}{c} v \cos \psi _m 
\end{equation}
where $v$ and $c$ are speed of the mobile user and the speed of light, respectively, $f _c$ is the carrier frequency of the communication, and $\psi _{m}$ is the incident radio wave angle with respect to the motion of the mobile user. Fig.~\ref{fig:multipathModel} illustrates relationship of a base station, a mobile user, and scatterers in a multipath mobile communication that suffers from fading.

Observed channel $c (t)$ in an actual communication can be decomposed into multiple path components $c _m(t)$ in the frequency domain based on this model. Different path components with different incident angles $\psi _m$ appear as separated peaks in a Doppler frequency spectrum. Hence, the parameters of each path component can be estimated by finding peak amplitudes and Doppler frequencies for $a _{m}$ and $f _{m}$ in the Doppler spectrum and the corresponding phase shifts for $\phi _{m}$ in its phase spectrum. Chirp z-transform (CZT) with a Hann window provides low calculation cost and a smooth frequency-domain interpolation that is useful for an accurate estimation of the parameters in the region close to zero frequency \citep{Tan2009}. By sliding the Hann window on preceding channel states and by repeating the parameter estimation process, we can obtain separated path components at different time points. We focus on the fact that the separated channel states $c_{m}(t)$ have rotary locus in the complex plane and predict its change in time for obtaining the future channel by using CVNNs. 

\section{Conventional CVNN-Based Channel Prediction with a Pre-Defined Network Structure}
\label{sec:conventionalCVNN}

The changes in the separated channel components $c_{m}(t)$ can be predicted by ML-CVNNs \citep{Ding2014b}. CVNN is a framework suitable for treating signal rotation and scale in the complex plane by use of its high generalization ability \citep{Hirose2012,Hirose2012a,Trabelsi2018}. It has been receiving more attention in various applications that intrinsically require dealing with complex values \citep{Hara2004,Kawata2005,Valle2014,Arima2017}. With a basic ML-CVNN consisting of a layer of $I_{\rm ML}$ input terminals, a hidden-neuron layer with $K_{\rm ML}$ neurons and an output neuron, we can predict the complex-valued $c_{m}(t)$ from a set of past channel components, $ {c} _{m} ( t-1 )$, ..., ${c} _{m} ( t-I_{\rm ML} )$ for each path $m = 1$, ..., $M$. The input terminals distribute input signals, $ {c} _{m} ( t-1 )$, ..., ${c} _{m} ( t-I_{\rm ML} )$, to the hidden-layer neurons as their inputs $\boldsymbol{z}_1 $. In the same way, the outputs of the hidden-layer neurons $\mbox{\boldmath $ z$} _2$ are passed to the output-layer neuron as its inputs. The neurons in the hidden layer are fully connected with the input terminals and the output-layer neuron. The output of the output-layer neuron $z _3$ is the prediction result $\tilde{c} _{m} (t)$. The connection weight $ w_{lkj}$ to $k$th output of $j$th neuron/input terminal in layer $l$ is expressed by its amplitude $| w_{lkj} |$ and phase $\theta _{lkj}$. The internal state $u _{(l+1)k}$ of $k$th neuron in $(l+1)$th layer is obtained as the summation of its inputs $\boldsymbol{z}_{l}$ weighted by $\boldsymbol{w}_{lk}=[w _{lkj}]$, i.e.,
\begin{equation}\label{eq:internalState}
u _{(l+1)k} = |u _{(l+1)k}|e^{i\theta_{(l+1)k}} \equiv \sum _j |w _{lkj}| |z_{lj}| e^{i (\theta_{lkj} + \theta_{lj})}
\end{equation}
where $z_{lj} = |z_{lj}| e^{j \theta_{lj}}$. The output $z_{(l+1)k}$ is given by adopting an amplitude-phase-type activation function $f_{\rm ap}$ to $u _{(l+1)k}$ as
\begin{equation}\label{eq:neuronOutput}
z_{(l+1)k} \equiv f_{\rm ap}(u_{(l+1)k}) = \tanh{(|u _{(l+1)k}|)} e ^{(i \theta _{(l+1)k})} 
\end{equation}

In our previous work, the connection weights $\boldsymbol{W} _{l}=[w _{lkj}]$ in the ML-CVNN were updated as follows. The ML-CVNN regarded the past known channel component $\hat{c} _{m} (t)$ as an output teacher signal, while the preceding channel components associated with the same path $\hat{c} _{m} (t-1), ..., \hat{c} _{m} (t-I_{\rm ML})$ were considered as input teacher signals. The weights have been updated based on the steepest descent method so that they minimize the difference 
\begin{equation}\label{eq:objectiveConv}
E _{(l+1)} \equiv \frac{1}{2}| \mbox{\boldmath $z$} _{(l+1)} - \hat {\mbox{\boldmath $z$}} _{(l+1)} |^{2}
\end{equation}
where $ \mbox{\boldmath $z$} _{(l+1)}$ and $\hat {\mbox{\boldmath $z$}} _{(l+1)}$ denote temporary output signals and the teacher signals, respectively, in layer $(l+1)$. The teacher signals in the hidden layer $\hat {\mbox{\boldmath $z$}} _{2}$ were the signals obtained through the backpropagation of the teacher signal (BPTS) of the output layer $\hat z_{3}$ \citep{Hirose1994,Hirose1996,Hirose2012}. The weight updates were performed at each estimated channel components by sliding the teacher signal and the input set in the time domain. We have stopped the update at a certain small number of iteration $R_{\rm ML}$ in the update process for $\hat{c} _{m}(t)$ and kept the updated weights as the initial values in the following weight update for $\hat{c} _{m}(t+1)$. With this procedure, we reduced the learning cost and followed the weak regularity of the separated channel components $c _{m}(t)$ for achieving a channel prediction with high accuracy.

\section{Proposal of Online Self-Optimizing CVNN}
\label{sec:dynamicCVNN}

A number of previous studies have proposed different methods to finding optimized structures of neural networks in general \citep{Ishikawa1996,Zou2005,Tzyy-ChyangLu2013,Ramachandram2017}. The so-called destructive neural networks start learning with a large structure, and then prune redundant connection weights and/or neurons to obtain an optimum network \citep{Karnin1990,Reed1993}, whereas the constructive neural networks raise the size from a small network to larger ones \citep{Elman1993,Barakat2011}.
 
\begin{figure}[!t]
\centering
\includegraphics[width=0.8\hsize]{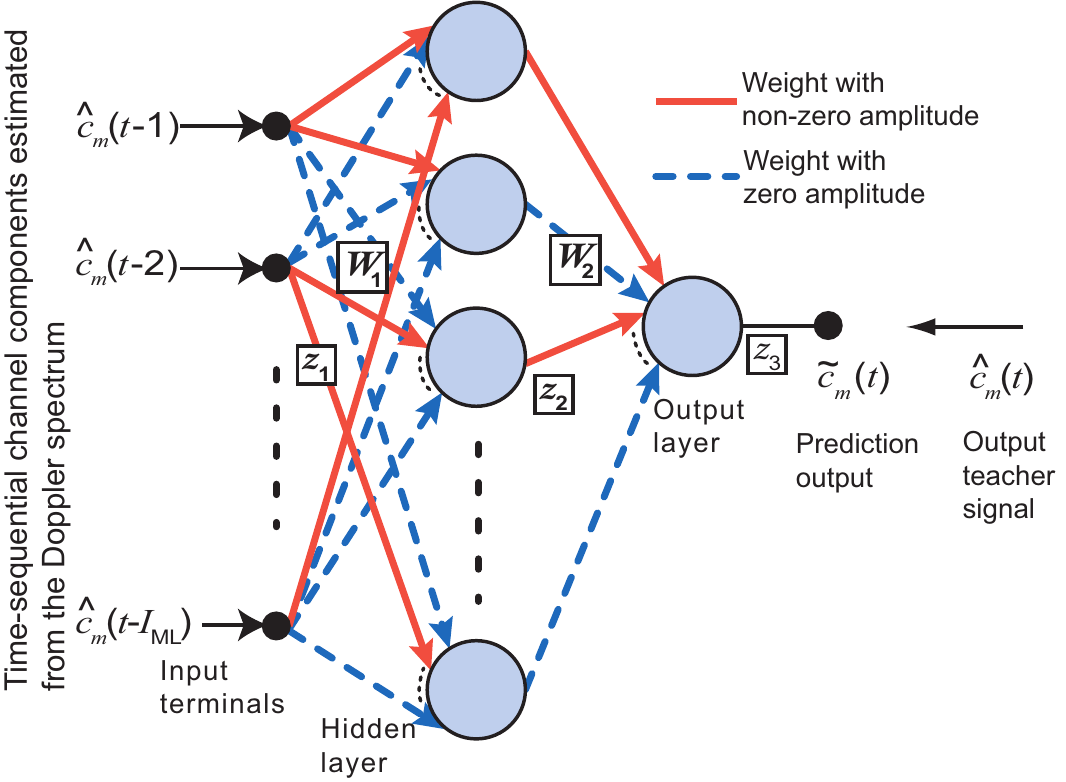}
\caption{Construction of the sparse complex-valued neural network, in which the solid arrows show non-zero-amplitude connections while dashed arrows represent zero-amplitude ones.}
\label{fig:neuralModelSparse}
\end{figure}

In this paper, we introduce regularization in the complex domain to achieve a dynamic CVNN that prunes and grows connections depending on the fluctuating communication situations. Fig.~\ref{fig:neuralModelSparse} shows the construction of the CVNN. We want a CVNN that changes its connections according to prediction situations and dynamically keeps suitable network structures in a series of predictions without manual tuning. To realize such a network, we introduce a constraint for sparsity to the weight updates in order to restrict the connections of networks to a suitably small size. The $L_{0}$-norm is an exact sparsity measure, and our problem can be redefined as minimizing the error function of the weights (\ref{eq:objectiveConv}) with the $L_{0}$-norm constraint on the connection weights. However, this problem has been shown to be NP-hard in general. Fortunately, under some conditions, the $L_{1}$-norm can serve as a sparsity measure for substituting the $L_{0}$-norm \citep{Tibshirani1996,Donoho2003,Gribonval2003}. The $L_{1}$-norm of the weights is a practical sparsity measure since it is convex so that we can perform optimization more easily \citep{Candes2006,Donoho2008,Elad2010}. By introducing the sparse constraint as a penalty function in a ML-CVNN, the objective function we use to update the weights in layer $l$ is expressed as
\begin{equation}\label{eq:objectiveSparse}
\arg \min_{\mbox{\boldmath $ W $}_{l}} E ^{\rm S} _{(l+1)} = \arg \min_{\mbox{\boldmath $ W $}_{l}} (\frac{1}{2} | \mbox{\boldmath $ z$}_{(l+1)} - \hat{\mbox{\boldmath $ z$}}_{(l+1)}|^2 + \alpha \| \mbox{\boldmath $ W $}_{l} \|_{1})
\end{equation}
where $\alpha$ is a coefficient to express degrees of the penalty. Minimizing the second term of the right-hand side of (\ref{eq:objectiveSparse}) means restricting the non-zero weight number to get its minimal number in the network. This is effectively equivalent to pruning connection weights. In other words, the penalty function introduces sparsity to the weight updates so that the remaining weights form an effective structure for representing the output signal. We use the steepest descent method in the complex domain to update the weights here (Appendix \ref{appendix:derivCompVSteepDescent}). Thus, the weight amplitude $| w _{lki} |$ and the phase $\theta _{lki}$ are renewed as
\begin{align}\label{eq:weightRenewAmpSparse}
| w _{lkj}  |  (r+1) = &| w _{lkj} | (r) - \kappa_1 \frac{\partial E ^{\rm S} _{(l+1)}}{\partial {(|w _{lkj}|)}} \nonumber \\ 
	= &| w _{lkj} | (r) - \kappa_1 \Big\{ ( 1- | z_{(l+1)k} | ^2 ) \nonumber \\   	
    &\times \big( | z_{(l+1)k} | -  | \hat{z}_{(l+1)k} | \cos{( \theta_{(l+1)k} - \hat{\theta} _{(l+1)k} )} \big) | z_{lj} | \cos{\theta^{\rm rot} _{lkj}} \\
    &- | z _{(l+1)k} | | \hat{z} _{(l+1)k} | \sin{ ( \theta _{(l+1)k} - \hat{\theta} _{(l+1)k} )} \frac{| z _{lj} |}{| u_{(l+1)k} |}  \sin{ \theta^{\rm rot} _{lkj}} \nonumber \\ 
    &+ \alpha \Big\} \nonumber
\end{align}
\begin{align}\label{eq:weightRenewPhaSparse}
\theta _{lkj}  (r + 1) = &\theta _{lkj}  (r) - \kappa_2 \frac{1}{| w _{lkj}|} \frac{\partial E ^{\rm S} _{(l+1)}}{\partial {\theta _{ lkj }}} \nonumber \\ 
	= & \theta _{lkj}  (r) - \kappa_2 \Big\{ ( 1- | z_{(l+1)k} | ^2 ) \nonumber \\
    & \times \big( | z_{(l+1)k} | -  | \hat{z}_{(l+1)k} | \cos{( \theta_{(l+1)k} - \hat{\theta} _{(l+1)k} )} \big) | z_{lj} | \sin{ \theta^{\rm rot} _{lkj}} \\
    &+ | z _{(l+1)k} | | \hat{z} _{(l+1)k} | \sin{( \theta _{(l+1)k} - \hat{\theta} _{(l+1)k} )} \frac{| z _{lj} |}{| u_{(l+1)k} |}  \cos{\theta^{\rm rot} _{lkj}} \Big\} \nonumber
\end{align}
where $\theta^{\rm rot} _{lkj} \equiv \theta _{(l+1)k} - \theta _{lj} - \theta _{lkj}$, $r$ is an index of learning iteration, and $\kappa _1$ and $\kappa _2$ are learning constants. This update rule has an additional term $+ \alpha$ in the amplitude $| w _{lki} |$ update in comparison to the conventional complex-valued steepest descent method \citep{Hirose2012,Ding2014b} because of the penalty term. 

In addition to the $L_{1}$-norm regularization, we also introduce group sparse term \citep{Yuan2006,Scardapane2017,Wang2018} as a penalty of the weight updates in this paper. While $L_{1}$-norm imposes sparsity on weight connections by considering each weight as a single unit, the group sparse penalty introduced by $L_{2,1}$-norm regularization imposes sparsity on input terminals and neurons in a network by considering all outgoing wights from a input terminal or a neuron as a single group. We prune input terminals and/or neurons by redefining the objective function as
\begin{align}\label{eq:objectiveGroupSparse}
\arg \min_{\mbox{\boldmath $ W $}_{l}} E^{\rm GS} _{(l+1)} &= \arg \min_{\mbox{\boldmath $ W $}_{l}} (\frac{1}{2} | \mbox{\boldmath $ z$}_{(l+1)} - \hat{\mbox{\boldmath $ z$}}_{(l+1)}|^2 + \alpha \| \mbox{\boldmath $ W $}_{l} \|_{2,1}) \nonumber\\
&= \arg \min_{\mbox{\boldmath $ W $}_{l}} \Big(\frac{1}{2} | \mbox{\boldmath $ z$}_{(l+1)} - \hat{\mbox{\boldmath $ z$}}_{(l+1)}|^2 + \alpha \sum _{j} \sqrt{|\boldsymbol{w}_{lj}|} \sqrt{\sum _{k} |w _{lkj}|^2} \Big)
\end{align}
where $\boldsymbol{w}_{lj} = [w _{lkj}]$ and $|\boldsymbol{w}_{lj}|$ denotes the dimensionality of the vector $\boldsymbol{w}_{lj}$. Note that, in this work, $\sqrt{|\boldsymbol{w}_{lj}|}$ can be moved outside the summation of $j$ for consisting isotropic $L_{2,1}$-norm due to the same dimensionality of weights in a layer resulting from the fully connected structure. Minimizing the $L_{2,1}$-norm term means restricting the input terminal/neuron numbers by setting all weight connections from a terminal or a neuron to be either simultaneously zeros or none of them are. We use the same steepest descent method to update the weights based on this objective (Appendix \ref{appendix:derivCompVSteepDescent}) and get a new update rule for the complex-valued group sparse as
\begin{align}\label{eq:weightRenewAmpGroupSparse}
| w _{lkj}  |  (r+1) = &| w _{lkj} | (r) - \kappa_1 \frac{\partial E ^{\rm GS} _{(l+1)}}{\partial {(|w _{lkj}|)}} \nonumber \\
	= &| w _{lkj} | (r) - \kappa_1 \Big\{ ( 1- | z_{(l+1)k} | ^2 ) \nonumber \\   	
    &\times \big( | z_{(l+1)k} | -  | \hat{z}_{(l+1)k} | \cos{( \theta_{(l+1)k} - \hat{\theta} _{(l+1)k} )} \big) | z_{lj} | \cos{\theta^{\rm rot} _{lkj}} \\
    &- | z _{(l+1)k} | | \hat{z} _{(l+1)k} | \sin{( \theta _{(l+1)k} - \hat{\theta} _{(l+1)k} )} \frac{| z _{lj} |}{| u_{(l+1)k} |}  \sin{ \theta^{\rm rot} _{lkj}} \nonumber \\ 
    &+ \alpha \sqrt{|\boldsymbol{w}_{lj}|} \frac{|w _{lkj}|}{\| \boldsymbol{w}_{lj} \|_2} \Big\} \nonumber
\end{align}
\begin{align}\label{eq:weightRenewPhaGroupSparse}
\theta _{lkj}  (r + 1) = &\theta _{lkj}  (r) - \kappa_2 \frac{1}{| w _{lkj}|} \frac{\partial E ^{\rm GS} _{(l+1)}}{\partial {\theta _{ lkj }}} \nonumber \\ 
	= & \theta _{lkj}  (r) - \kappa_2 \Big\{ ( 1- | z_{(l+1)k} | ^2 ) \nonumber \\
    & \times \big( | z_{(l+1)k} | -  | \hat{z}_{(l+1)k} | \cos{( \theta_{(l+1)k} - \hat{\theta} _{(l+1)k} )} \big) | z_{lj} | \sin{\theta^{\rm rot} _{lkj}} \\
    &+ | z _{(l+1)k} | | \hat{z} _{(l+1)k} | \sin{ ( \theta _{(l+1)k} - \hat{\theta} _{(l+1)k} )} \frac{| z _{lj} |}{| u_{(l+1)k} |}  \cos{\theta^{\rm rot} _{lkj}} \Big\} \nonumber
\end{align}
This update rule has an additional term $+ \alpha \sqrt{|\boldsymbol{w}_{lj}|} |w _{lkj}| / \| \boldsymbol{w}_{lj} \|_2$ in the amplitude $| w _{lki} |$ update in comparison to the conventional complex-valued steepest descent method due to the $L _{2,1}$-norm penalty.

For simplicity and lower computational consumption, the BPTS is used in the both update schemes for getting the teacher signal $\hat{\mbox{\boldmath $ z$}}_2 $ in the hidden layer from the teacher signal in the output layer $\hat z_{3}$ as
\begin{equation}\label{eq:bpts}
\hat{\mbox{\boldmath $z$}} _{2} = 
\left( 
f_{\rm ap}(\hat{z} ^{*} _{3} \boldsymbol{W} _{2}) 
\right) ^{*}
\end{equation}
where $( \cdot )^{*}$ represents the complex conjugate or hermite conjugate.

\begin{table}[t]\centering
 \caption{Communication Parameters}
 \label{tbl:commParameters}
 \begin{tabular}{ l | l }\hline
  Parameter & Value  \\ \hline \hline 
  QPSK symbol number & 12852 \\ 
  Number of OFDM subcarriers & 2048 \\ 
  Number of OFDM guard bands & 106 left, 106 right \\ 
  Number of OFDM symbols & 7 \\ 
  Length of OFDM cyclic prefix & [160, 144$\times$6] \\ 
  TDD sub-frame length & 0.512 ms \\ 
  TDD symbol number in a sub-frame & 15360 \\
  TDD frame length & 10 sub-frames \\
  Sampling rate & 30 MHz \\ \hline 
 \end{tabular}
\end{table}

To predict fading channels, we update the connection weights by time-sliding the inputs and output teacher signals on the estimated $\hat{c} _{m}(t)$ sequences as we performed in the previous work \citep{Ding2014b}. That is, a set of updated weights using the complex-valued estimation $\hat{c} _{m}(t)$ as the output teacher signal for $\tilde{c} _{m}(t)$ in Fig.~\ref{fig:neuralModelSparse} and $\hat{c} _{m} (t-1), ..., \hat{c} _{m} (t-I_{\rm ML})$ as the input signals are kept in the network and used as the initial weights in the following update for $\tilde{c} _{m}(t+1)$ by regarding $\hat{c} _{m}(t+1)$ as the new output teacher signal and $\hat{c} _{m} (t), ..., \hat{c} _{m} (t-I_{\rm ML}+1)$ as the new input signals. The weight update is performed until the latest channel component is used and the most up-to-dated weight connections predict the future channel states. The combinations of the penalty terms and the prediction scheme in the time domain are expected to keep the structure to be a suitable size for the channel prediction depending on the fluctuating communication environment.

The computation complexity of the online training part is $O(M \cdot R _{\rm ML} \cdot K _{\rm ML} \cdot I _{ML})$ where $M$, $R _{ML}$, $K _{ML}$, and $I _{ML}$ are the detected path number, the weight update iterations, the hidden-neuron number and the input terminal number, respectively. This complexity is equivalent to that of the conventional ML-CVNN updates \citep{Ding2014b} and typically much smaller than the complexity of CZT calculation, $O(N{\rm log}N)$ where $N$ is the symbol number in a CZT window. The additional penalty functions based on $L _1$-norm and $L _{2,1}$-norm realize the per-weight and per-neuron sparsity within a CVNN structure without increase of calculation complexity for the weight updates.

\section{Numerical Experiments}
\label{sec:simulation}
In the following two sections, we evaluate the performance of the channel prediction methods based on the ML-CVNNs with the penalties in simulations and experiments. We assume orthogonal frequency-division multiplexing (OFDM) with quadrature phase shift keying (QPSK) modulation and time division duplex (TDD) as the communication scheme in this paper. For the future compatibility with 5G communications, cyclic-prefix (CP) OFDM with the system parameters listed on Table~\ref{tbl:commParameters} are used in this work. 

\begin{figure}[!t]
\centering
\includegraphics[width=3.33in]{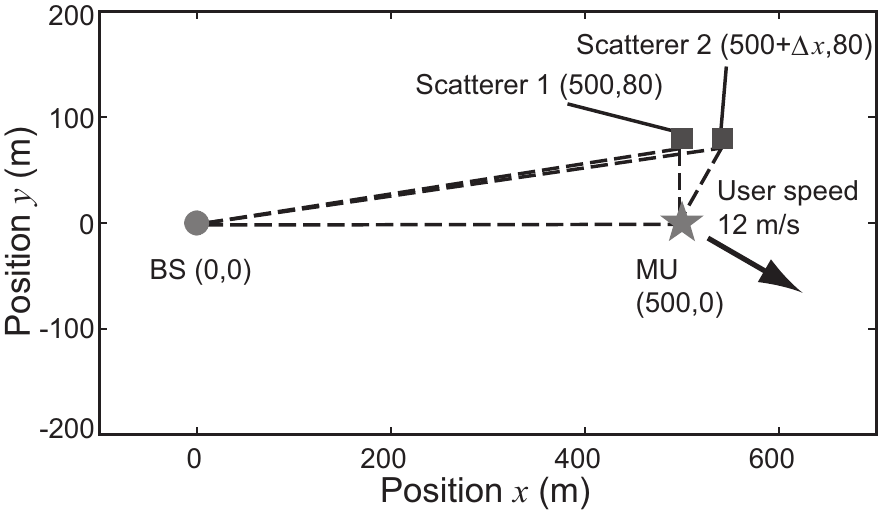}
\caption{Geometrical setup used in the simulation. There are two scatterers separated by $\Delta x$~m, a base station (BS) and a mobile user (MU) in an open communication space. The line of sight between the BS and the MU is considered. The MU moves in the direction of the arrow ($-30^{\circ}$ from the $x$ axis) with a velocity of 12~m/s.}
\label{fig:simGeometric}
\end{figure}

\begin{table}[t]\centering
 \caption{Channel Prediction Parameters}
 \label{tbl:channelPredictionParameters}
 \begin{tabular}{ l | l }\hline
  Parameter & Value  \\ \hline \hline 
  CZT size & 8 TDD frames \\ 
  Down-sampled signal rate for CZT & 500 kHz \\ 
  ML-CVNN input terminals $I_{\rm ML}$ & 30 \\ 
  ML-CVNN hidden-neuron number $K_{\rm ML}$ & 30 \\ 
  ML-CVNN weight update iterations $R _{\rm ML}$ & 10 \\ \hline 
 \end{tabular}
\end{table}

In this section, we characterize the performance of the proposed channel prediction methods with various degree of penalty $\alpha$ by using simulated fading channels. The geometrical setup of the simulation is shown in Fig.~\ref{fig:simGeometric}. We consider communications between a base station (BS) and a mobile user (MU) moving away from the BS at $12$~m/s with a certain moving angle. There are two scatterers making 2~paths in addition to the line-of-sight path. The carrier frequency is 2~GHz here.

We predict channel changes in a TDD frame based on its preceding channel states. The past path characteristics are estimated by using CZT with the Hann window. A window with 8-TDD-frame length is applied to the past channel states for estimating the path parameters, $a _m (t)$, $f_m (t)$, $\phi _m (t)$, based on peaks in Doppler spectra and corresponding phase spectra. Then, the past path characteristics $c _m (t)$ are composed by using the parameters and assigned as the estimated characteristics at the center of the window. We shift the window center at a TDD-frame interval for estimating multipath characteristics at every TDD frame. The details of the time frames are explained in our previous work \citep{Ding2014b}.


\begin{figure}[!t]
\centering
\includegraphics[width=2.9in]{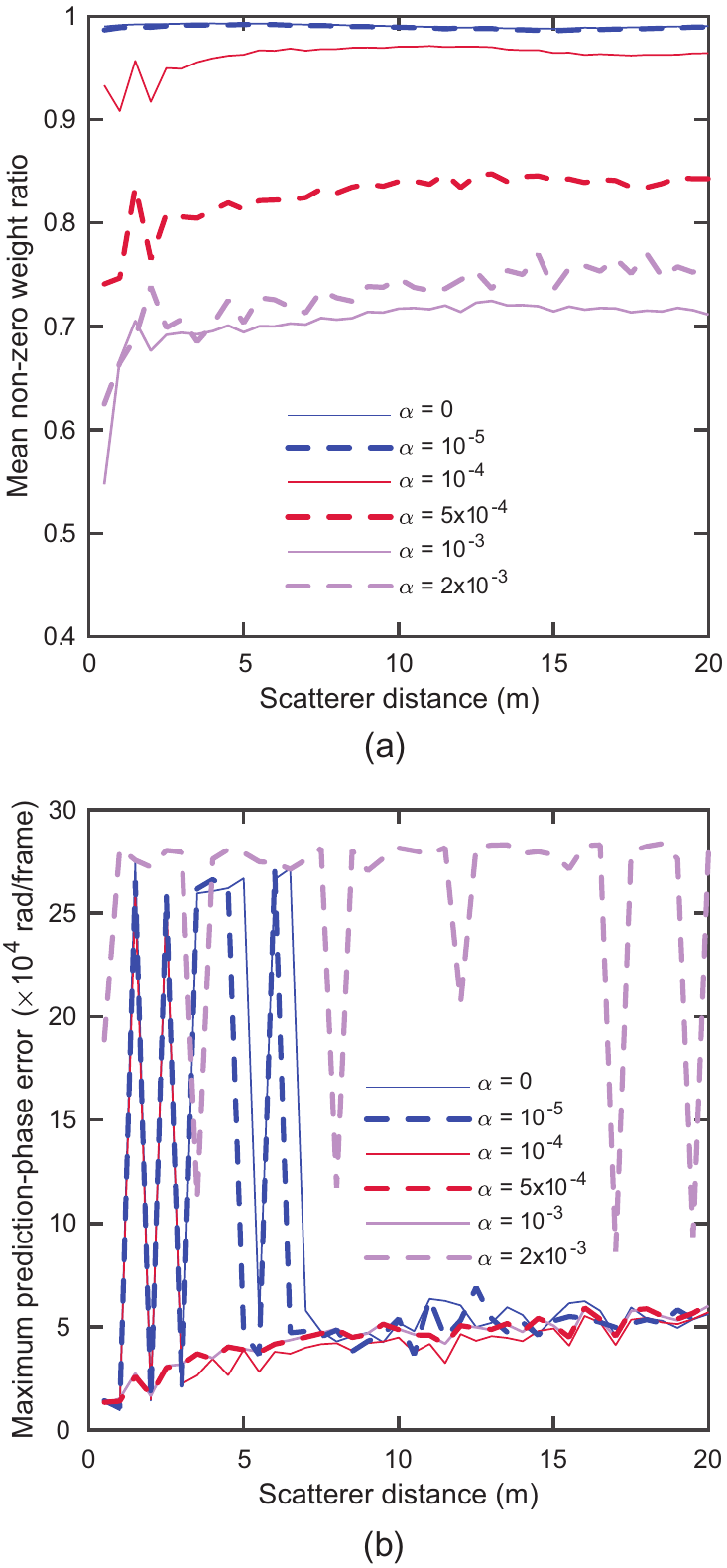}
\caption{Simulation results of the ML-CVNN-based channel prediction with the $L_1$-norm penalty when varying penalty coefficients $\alpha$. (a) Averaged non-zero weight ratios (network size) and (b) maximum predicted phase errors (prediction stability) against scatterer distance $\Delta x$ in Fig.~\ref{fig:simGeometric} (communication situations).}
\label{fig:alphaCompSim}
\end{figure}

\begin{figure}[!t]
\centering
\includegraphics[width=2.9in]{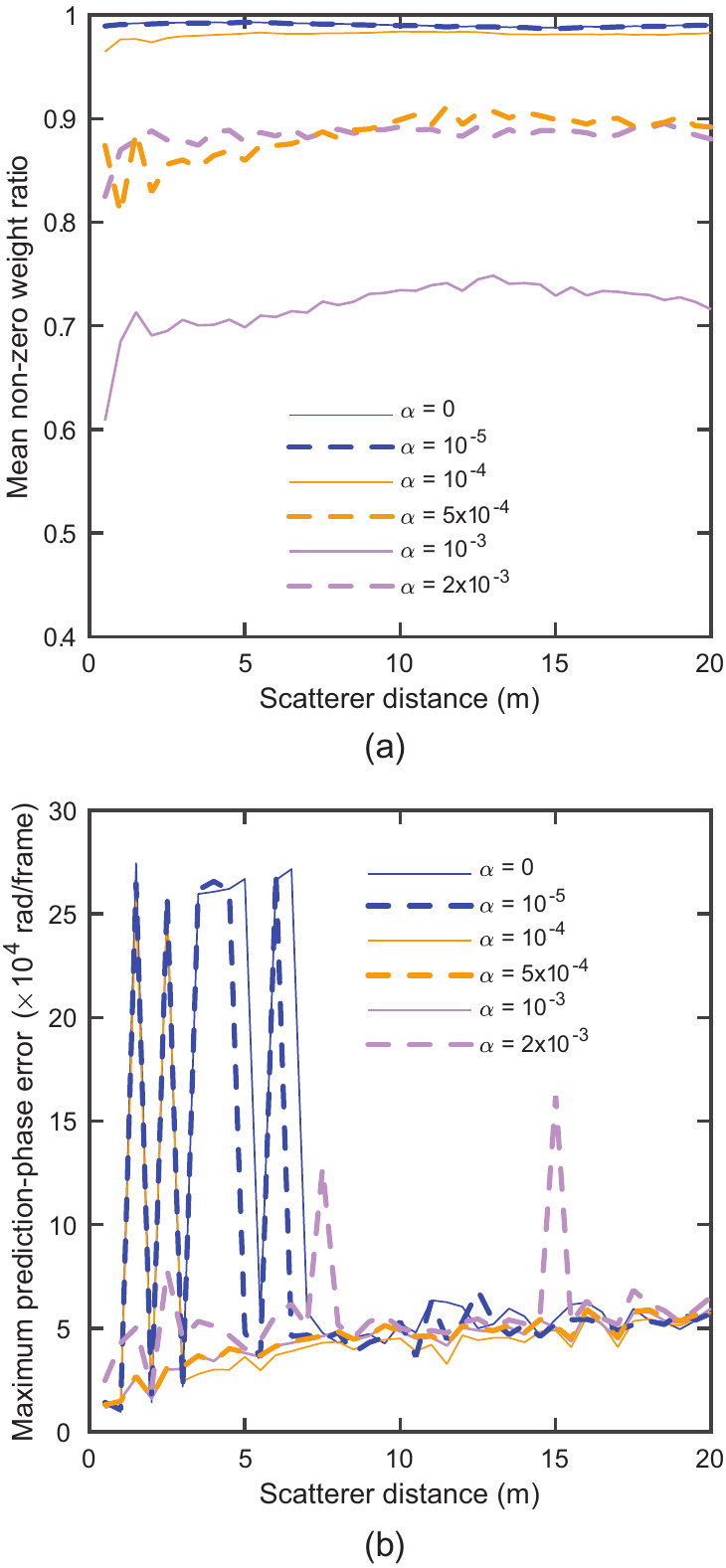}
\caption{Simulation results of the ML-CVNN-based channel prediction with the $L_{2,1}$-norm penalty when varying penalty coefficients $\alpha$. (a) Averaged non-zero weight ratios (network size) and (b) maximum predicted phase errors (prediction stability) against scatterer distance $\Delta x$ in Fig.~\ref{fig:simGeometric} (communication situations).}
\label{fig:alphaCompSimGroup}
\end{figure}

To evaluate the performance in various channel changes, we changed the scatterer distance $\Delta x$ shown in Fig.~\ref{fig:simGeometric} from $\Delta x = 0.5$ to $20$ m with $0.5$ m step, and performed $100$ independent predictions at different time points along the movement of MU in each scatterer arrangement. We started with the neural network with the parameters listed in Table~\ref{tbl:channelPredictionParameters}. The penalties prune and grow the network connections, $30 \times 30$ in the hidden layer and $30 \times 1$ in the output layer, in a timely manner as the communication situation changes.

Figs.~\ref{fig:alphaCompSim}(a)~and~\ref{fig:alphaCompSimGroup}(a) show the mean of the network size versus scatterer distance for the methods with the $L_1$-norm and the $L_{2,1}$-norm penalties, respectively. A connection weight is counted as non-zero here if its amplitude satisfies
\begin{equation}\label{eq:nonZeroWeight}
|w _{lkj}| \geq \max (|\boldsymbol{W} _{l}|)/100
\end{equation}
Otherwise, the weight is considered as a zero weight. If a weight in the output layer ($l = 2$) is counted as a zero weight, all the weights in the hidden layer connecting themselves to the corresponding neuron are also considered as zero weights in order to fairly compare the penalty effect on the entire network. The network sizes of the ML-CVNN with various penalty coefficients ($\alpha = 0, 10^{-5}, 10^{-4}, 5\times10^{-4}, 10^{-3}, 2\times10^{-2}$) have been evaluated, and the mean connection numbers for the 100 trials in each condition have been normalized by the maximum possible connections to show the non-zero connection ratio. Corresponding prediction accuracy is also calculated by accumulating predicted phase errors within the prediction frame. Figs.~\ref{fig:alphaCompSim}(b)~and~\ref{fig:alphaCompSimGroup}(b) present the maximum estimated phase errors in each communication condition out of the 100 predictions, showing stability of the predictions with the $L_1$-norm and the $L_{2,1}$-norm penalties, respectively.

We found in Figs.~\ref{fig:alphaCompSim}(a)~and~\ref{fig:alphaCompSimGroup}(a) that the non-zero weight number consisting an effective network decreases as the penalty coefficient $\alpha$ increases as expected, whereas a network without the penalties ($\alpha = 0$) keeps almost all of the connections active regardless the change of the communication environment. In Fig.~\ref{fig:alphaCompSim}(b)~and~\ref{fig:alphaCompSimGroup}(b), the smaller networks achieved by the penalties show better prediction stability compared to the conventional ML-CVNN-based method without any constraint ($\alpha = 0$). The results also present that the proposed prediction methods reaches its best performance with a penalty coefficient around $\alpha = 5\times10^{-4} \sim 10^{-3}$, and that $\alpha$ larger than this value introduces instability to the channel prediction again. Note that, the $L_{1}$-norm penalty tends to prune more connection weights within the networks than the $L_{2,1}$-norm does for the penalty coefficient $\alpha \leq 10^{-3}$ and for almost all of the scatterer arrangements we evaluated. Interestingly, Figs.~\ref{fig:alphaCompSim}~and~\ref{fig:alphaCompSimGroup} also depict that the prediction stability of different coefficients, i.e., $\alpha = 5\times10^{-4}$ and $\alpha = 10^{-3}$, only show little difference for both the regularization methods even though the non-zero weight ratios of those are different from each other. This fact demonstrates that the differently connected networks generated by the $L_1$-norm and $L_{2,1}$-norm penalized CVNNs provide comparable prediction performance although the internal realizations are different. Finally, these results show that the proposed prediction methods with an appropriate $\alpha$ can automatically prune redundant connections in its network to achieve higher prediction accuracy even in prediction conditions that are difficult for the conventional method. 

\begin{figure}[!t]
\centering
\includegraphics[width=3.33in]{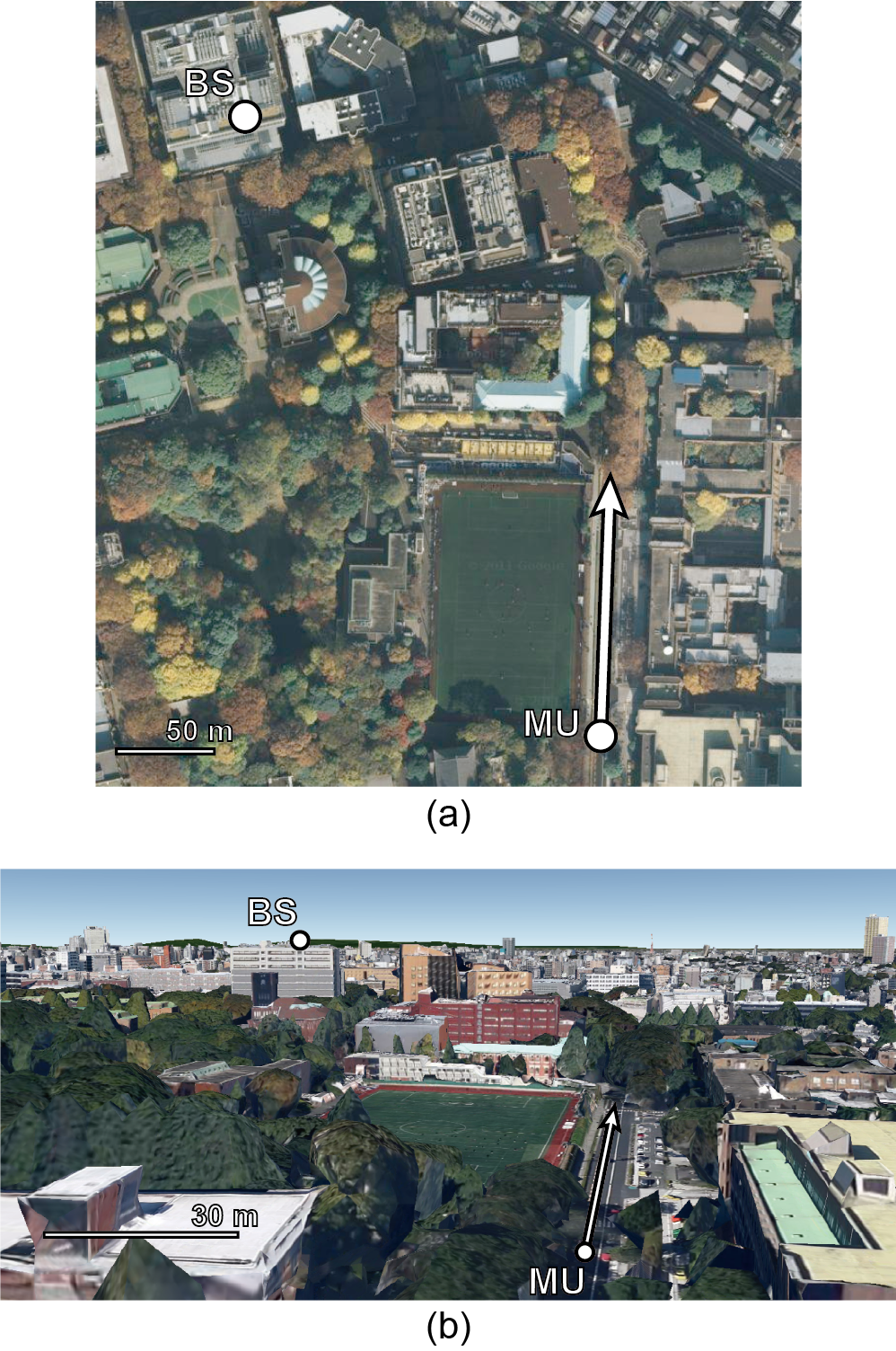}
\caption{Geometrical setup of the experiment illustrated as (a) two-dimensional top view (Google Maps, modified) and (b) three-dimensional side view (Google Earth, modified) which includes a fixed base station (BS), a moving mobile user (MU) and other obstacles.}
\label{fig:experimentEnvironment}
\end{figure}

\section{Experiments in Actual Communication Environment}
\label{sec:experiment}

\begin{figure}[!t]
\centering
\includegraphics[width=4.7in]{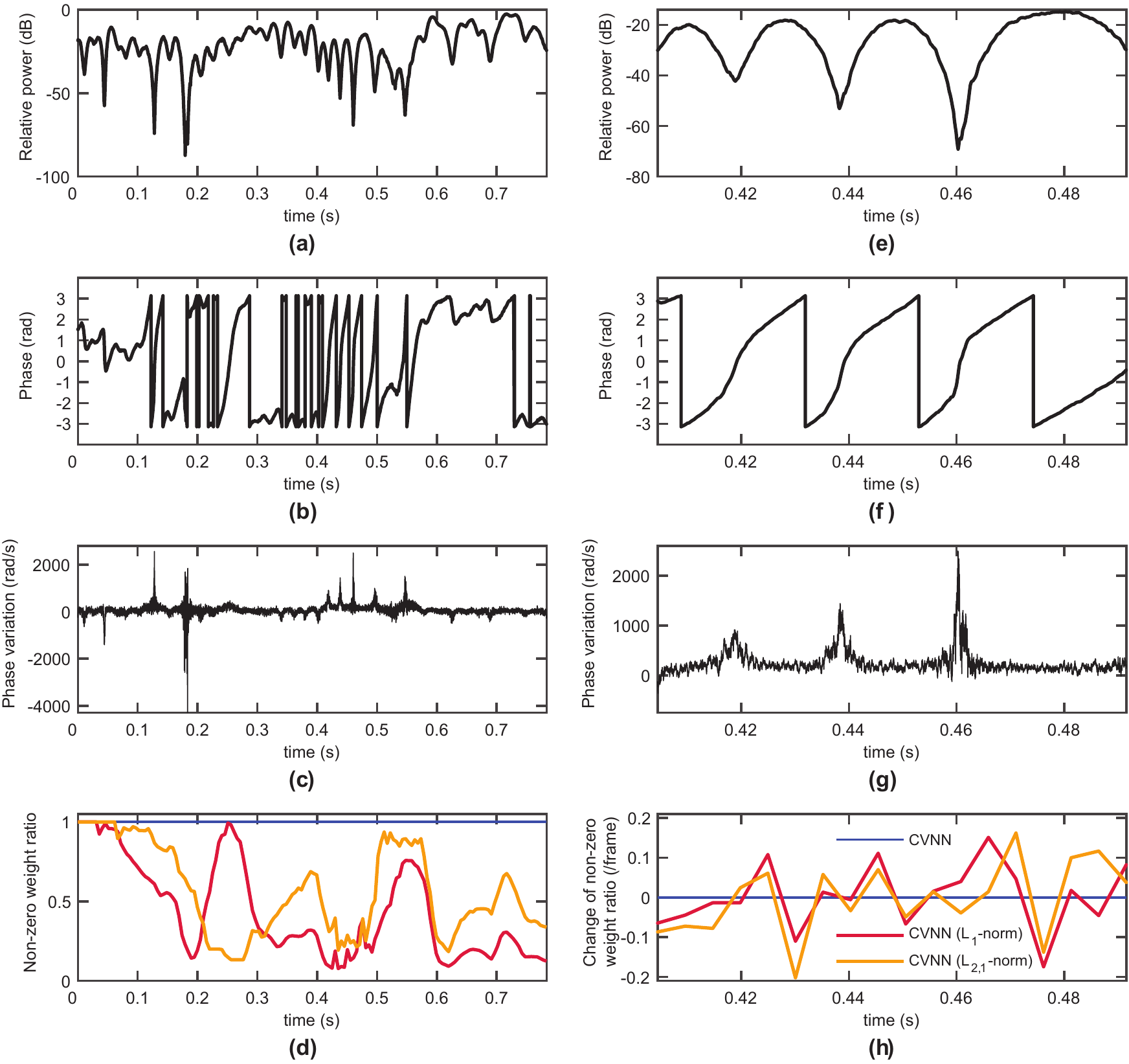}
\caption{Actual propagation-experiment results showing (a) observed relative power, (b) observed phase value, (c) change rate of the phase in every $2\times10^{-6}$~ms, and (d) non-zero weight ratios of CVNNs with $L_{1}$-norm (red) and $L_{2,1}$-norm (orange) penalties as well as without any penalty (blue). (e-g) zoomed-in version of a-c from 0.405 to 0.492 ms. (h) change rates of non-zero weight ratios of the CVNNs in this period. $\alpha = 5 \times 10^{-4}$ is assigned as the penalty degree for the proposed methods.}
\label{fig:structureCompExp}
\end{figure}

\begin{figure}[!t]
\centering
\includegraphics[width=3.33in]{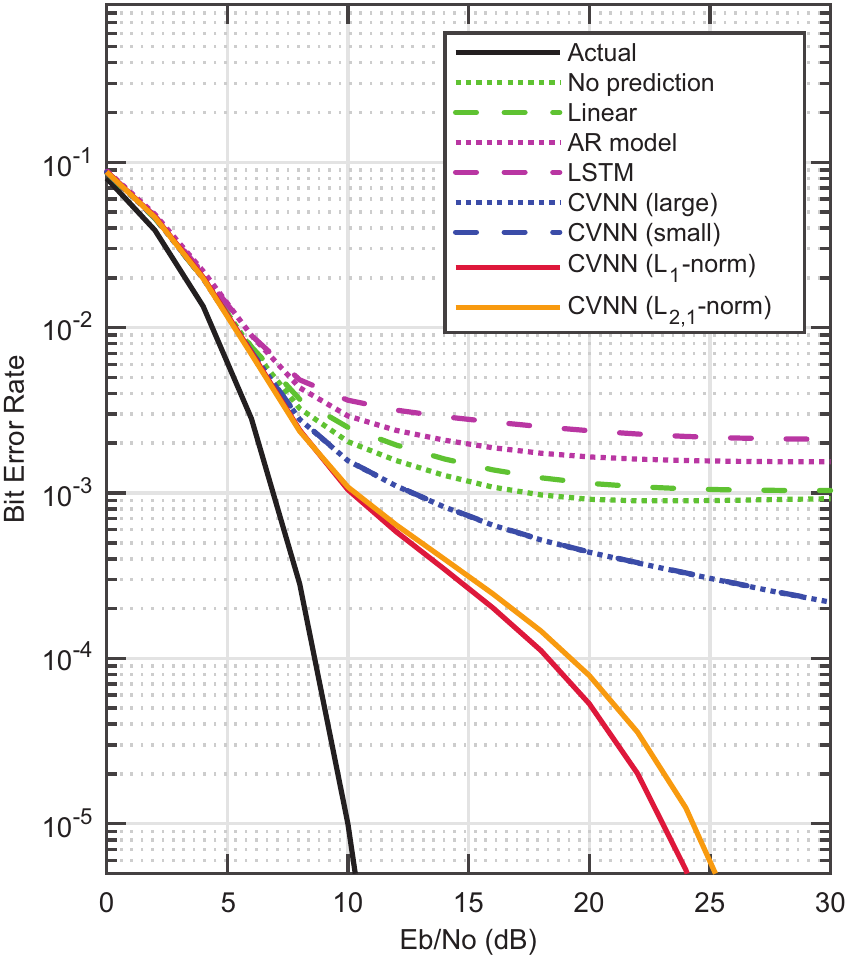}
\caption{BER curves obtained for different channel prediction methods in communications with fading channel measured in actual environment.}
\label{fig:BERSNRCompExp}
\end{figure}

In this section, we further demonstrate adaptability of the proposed methods in prediction with actually observed fading channels. We experimentally observed fading channels in a communication situation shown in Fig.~\ref{fig:experimentEnvironment}. There are an MU as a transmitter and a BS as a receiver in the experimental site with some obstacles, such as buildings, trees and other automobiles, providing a typical mobile communication environment in an urban area. The MU moves in the direction of the arrows shown in Figs.~\ref{fig:experimentEnvironment}(a) and (b) with a velocity around 12.5~m/s and transmits 1.297~GHz nonmodulated waves from a monopole antenna, whereas the BS receives the wave by using another monopole antenna. Note that, depending on communication instants, a line-of-sight path may or may not present due to obstacles. The received channel signal was mixed with a 1.287~GHz local oscillator wave after an amplifier, and then extracted as a signal at an intermediate frequency of 10~MHz. After passing it to another amplifier and a band-pass filter with 2~MHz bandwidth, we sample the channel information at 30~M~Sample/s. The channel was further down-sampled to 500~kHz for reduction of computational requirements in the CZT estimation. The observed channel change has already been shown in Fig.~\ref{fig:channelChangeExp} as an example of the fading captured in this communication situation. It is time-sequential data showing irregular rotation of the channel in the complex plane received at the BS. The channel state gives roughly 2 distinct main paths over the communication period. The channel changes in a TDD frame are predicted based on preceding channel states by using CZT and ML-CVNNs in the same way as in Section~\ref{sec:simulation}. 

First, we evaluate the time variation of the ML-CVNN size to demonstrate the online dynamics. Fig.~\ref{fig:structureCompExp} shows sequential changes of a fading channel and the neural-network size in a prediction process. The actually received signal power, phase values and change rates of the phase are shown in Figs.~\ref{fig:structureCompExp}(a-c), respectively. Non-zero connection weights are counted by using the same scheme as (\ref{eq:nonZeroWeight}) in Section~\ref{sec:simulation}, and plotted against time in Fig.~\ref{fig:structureCompExp}(d). In order to demonstrate the impact of the penalty functions on the network size change, $\alpha=5 \times 10^{-4}$ has been used based on the discussion in the previous section. For a comparison, the operation without any penalty constraint is also characterized as the conventional method. 

Fig.~\ref{fig:structureCompExp}(c) shows that the channel does not always change in the same manner but there are sporadic fast changes among relatively stable states. The fast changes cause difficulty in channel prediction and degrade its performance. In Fig.~\ref{fig:structureCompExp}(d), we can observe that the proposed methods with the penalty functions, namely CVNN ($L_{1}$-norm) and CVNN ($L_{2,1}$-norm), increase their effective structure sizes at and/or after the large channel changes while the entire trend of the size is kept to be relatively compact through out the process. On the other hand, the conventional method without any penalty, namely CVNN, does not change its network size in any part of the update procedure, and no correlation with the channel changes is observed.

For further discussion, we focus on a prediction period containing three fast phase jumps. Figs.~\ref{fig:structureCompExp}(e-g) present the channel power, phase and phase changes during the period from $405$~ms to $492$~ms in Figs.~~\ref{fig:structureCompExp}(a-c). Fig.~\ref{fig:structureCompExp}(h) shows the change rates of the non-zero weight ratio in the period calculated by taking the difference of the weight ratios at two consecutive TDD frames. It is obvious that the CVNNs with the regularization increase their non-zero weight connections during their update process synchronously with the large channel changes in order to adapt the network structures to those difficult prediction parts, and decrease the connections after them. In contrast, the conventional method is not sensitive to the channel changes. These results demonstrate that the CVNNs with the penalty functions accommodates themselves to such large and irregular channel changes by increasing their weight connections while they reduce the connections when the channel changes steadily. In other words, the proposed method has the ability to change the network structure dynamically and adaptively online according to the degree of difficulty in the channel prediction.

Finally, we compare prediction accuracy in various channel prediction methods in the actual communications. In this test, respective methods predict fading channel states in each TDD frame using channel information prior to the prediction periods in the same way as above. The predicted channel states are used for compensating the true fading in a communication situation with the CP-OFDM system described is Section~\ref{sec:simulation}. A randomly generated binary sequence has been converted into QPSK symbols and modulated into transmission signals based on the parameters shown in Table~\ref{tbl:commParameters}. The signals are assumed to be transmitted through a communication environment with the observed fading channel and different levels of additive white Gaussian noise. Before demodulation of OFDM and QPSK, the received signals are compensated by the predicted channel states using multiple prediction methods, namely, methods based on a linear prediction directly in the time domain, an AR model using channel characteristics estimated by CZT, a LSTM network \citep{Hochreiter1997,Cho2014} (Appendix \ref{appendix:lstmNet}), the conventional CVNN, and the proposed CVNN with the $L_1$-norm and $L_{2,1}$-norm penalties. We independently performed this process 101 times on different periods of the observed fading channels. For demonstrating the performance of the proposed methods, we also evaluate the conventional CVNN-based method with a smaller network size, namely CVNN (small), consisting of the same input terminals $I _{\rm ML} = 30$ but with smaller hidden-neuron number $K _{\rm ML} = 5$ in addition to the network structure listed in Table~\ref{tbl:channelPredictionParameters}.

Fig.~\ref{fig:BERSNRCompExp} shows the bit-error-rate (BER) curves against bit-energy to noise-power-density ratio $E_b$/$N_0$.  Here, CVNN (large) shows the result of the conventional CVNN-based method with the network structure listed in Table~\ref{tbl:channelPredictionParameters}, whereas CVNN ($L_1$-norm) and CVNN ($L_{2,1}$-norm) present that of the proposed methods with the same structure size and a penalty coefficient $\alpha = 5 \times 10 ^{-4}$. Communication without any prediction method, that is, channel compensation using channel states in the most recent TDD frame, is also performed for a comparison (No prediction). Communication bit errors with respective prediction methods are obtained in each process, and accumulated over all iterations for the final BER calculation. The black solid curve in the figure represents BER if the true future channel is perfectly known, thus showing the lower bound of the BER with the considered CP-OFDM setup. We can see the difficulty of the channel prediction on the actual fading in the large deviation of the BER curves corresponding to the linear, AR-model-based, LSTM and the conventional CVNN-based (for both large and small) prediction methods. The larger error rates by some of the methods, that is AR model, LSTM and CVNN (larger), compared to the no-prediction method implies that observed rapid and irregular changes of the channel states cause failure of channel compensation by those methods. In contrast, the proposed methods with the regularization, CVNN ($L_1$-norm) and CVNN ($L_{2,1}$-norm), achieve an accurate prediction even in the difficult communication situations and give $10^{-5}$ BER at $E_b$/$N_0$ $= 23-24 $ dB. The results show that the proposed online adaptive CVNNs with the regularization provides higher channel prediction performance. The clear improvement of the proposed methods compared to CVNN (large) and CVNN (small) shows the benefit of the online regularization methods over the stationary networks in the channel prediction.

\section{Conclusions}
\label{sec:conclusion}
In this paper, we proposed an online adaptive channel prediction method based on ML-CVNNs with self-optimizing dynamic structures. One of the main challenges in channel prediction with fast fading is its randomly time-varying channel states. For example, even a slight environment change due to movement of mobile users and/or obstacles during signal transmissions may significantly alter the received channel at the communication end. Hence, it is extremely difficult to construct a universal learning model covering such wide ranges of changing communication scenarios \citep{OShea2017}. Here, we believe that the combination of the shallow network and the regularization-based weight update methods with the online learning-prediction scheme can overcome these fundamental difficulties of the fast fading prediction, and provide practical and computationally non-expensive prediction with high prediction accuracy. Simulations and experiments demonstrated that the proposed CVNNs automatically change their effective connection numbers depending on the channel variation so that they keep appropriate network sizes to achieve accurate channel prediction without prior knowledge of the communication environments. The results presented in the experiment section for actually observed channels showed that the proposed method can provide more accurate prediction even in situations that are difficult for conventional methods including the time-domain linear, the AR-model-based, LSTM-based, and the conventional CVNN-based predictions.

\section*{Acknowledgment}
This work was supported in part by JSPS KAKENHI Grant No.~18H04105 and also by the Cooporative Research Project Program of the RIEC, Tohoku University. The authors are grateful to the anonymous reviewers for their insightful suggestions. 

\begin{appendix}
\section{Derivation of complex-valued steepest descent methods with $L_{1}$-norm and $L_{2,1}$-norm penalties}
\label{appendix:derivCompVSteepDescent}
We consider the complex-valued steepest descent methods with the sparse constraints ($L_{1}$- and $L_{2,1}$-norm penalties) in a multi-layered network structure. Here, we consider weight updates at connections from layer $l$ to layer $(l+1)$. The connection weight $w _{lkj}$ to $k$th output of $j$th neuron/input terminal in layer $l$ is expressed by its amplitude $|w _{lkj}|$ and phase $\theta _{lkj}$. Input signals to neurons in layer $(l+1)$ are output signals from layer $l$:
\begin{equation}
    z _{lj} = |z _{lj}|e ^{i\theta _{lj}}
\end{equation}
The internal state $u _{(l+1)k} = |u _{(l+1)k}| e ^{i \theta _{(l+1)k}}$ of $k$th neuron in $(l+1)$th layer is obtained as the summation of the inputs $\mbox{\boldmath $z$} _{l} = [z _{lj}]$ weighted by $\mbox{\boldmath $w$} _{lk} = [w _{lkj}]$, i.e.,
\begin{equation}
    u _{(l+1)k} \equiv \sum _j w _{lkj} z _{lj} = \sum _j |w _{lkj}| |z_{lj}| e^{i (\theta_{lkj} + \theta_{lj})}
\end{equation}
The output $z_{(l+1)k}$ of $k$th neuron in $(l+1)$th layer is, then, given by adopting an amplitude-phase-type activation function $f_{\rm ap}$ to $u _{(l+1)k}$ as
\begin{align}
z_{(l+1)k} \equiv & f_{\rm ap}(u_{(l+1)k}) \nonumber \\
=& {\rm tanh}(|u _{(l+1)k}|) e ^{i {\rm arg}(u _{(l+1)k})} = {\rm tanh}(|u _{(l+1)k}|) e ^{i \theta _{(l+1)k}}
\end{align}

Here, we use $dw _{lkj'}/dt$ to represent the descent direction of a weight $w _{lkj'}$ in $\mbox{\boldmath $w$} _{lk}$. Then, the change direction of the internal state $u _{(l+1)k}$ due to the weight change can be expressed as
\begin{align}\label{eq:internalStateDt}
\frac{du_ {(l+1)k}}{dt} = \Big( \frac{d(|u _{(l+1)k}|)}{dt} + i |u _{(l+1)k}| \frac{d\theta _{(l+1)k}}{dt} \Big) e ^{i\theta _{(l+1)k}}
\end{align}
If we newly define the descent direction of the weight as
\begin{align}\label{eq:weightChangeAP}
\frac{dw_ {lkj'}}{dt} \equiv \Big( \frac{dw ^{\rm a} _{lkj'}}{dt} + i \frac{dw ^{\rm p} _{lkj'}}{dt} \Big) e ^{i(\theta _{(l+1)k} - \theta _{lj'})}
\end{align}
by introducing two weight-change fractions in the directions of the weight amplitude and phase, $dw ^{\rm a} _{lkj'}/dt$ and $w ^{\rm p} _{lkj'}/dt$, on the complex plane, the change of the internal state $u _{(l+1)k}$ can also be represented as:
\begin{align}
\frac{du_ {(l+1)k}}{dt} = \frac{dw_ {lkj'}}{dt} z _{lj'} = \Big( \frac{dw ^{\rm a} _{lkj'}}{dt} + i \frac{dw ^{\rm p} _{lkj'}}{dt} \Big) |z _{lj'}| e ^{i\theta _{(l+1)k}}
\end{align}
Accordingly, we can obtain the change of the internal state $u _{(l+1)k}$ in terms of the weight-change fractions, for generalized $j = j'$, as
\begin{align}
\label{eq:dabsudwa} \frac{d|u_ {(l+1)k}|}{dw ^{\rm a} _{lkj}} =& |z _{lj}| \\
\label{eq:dangleudwp} \frac{d\theta _{(l+1)k}}{dw ^{\rm p} _{lkj}} =& \frac{|z _{lj}|}{|u _{(l+1)k}|} \end{align}

Here, we obtain the relationship between the change of the weight $dw_ {lkj'}/dt$ and the two fractions $dw ^{\rm a} _{lkj'}/dt$ and $w ^{\rm p} _{lkj'}/dt$ by dividing the both sides of (\ref{eq:weightChangeAP}) by $e ^{i\theta _{lkj'}}$ as
\begin{align}\label{eq:weightChangeDevAngleAP}
\frac{1}{e ^{i\theta _{lkj'}}} & \frac{dw_ {lkj'}}{dt} = \Big( \frac{dw ^{\rm a} _{lkj'}}{dt} + i \frac{dw ^{\rm p} _{lkj'}}{dt} \Big) e ^{i(\theta _{(l+1)k} - \theta _{lj'} - \theta _{lkj'})} \nonumber\\
 =& \cos{\theta ^{\rm rot} _{lkj'}} \frac{dw ^{\rm a} _{lkj'}}{dt} - \sin{ \theta ^{\rm rot} _{lkj'}} \frac{dw ^{\rm p} _{lkj'}}{dt} + i \Big( \cos{\theta ^{\rm rot} _{lkj'}} \frac{dw ^{\rm p} _{lkj'}}{dt} + \sin{ \theta ^{\rm rot} _{lkj'}} \frac{dw ^{\rm a} _{lkj'}}{dt}\Big)
\end{align}
where $\theta ^{\rm rot} _{lkj'} \equiv \theta _{(l+1)k} - \theta _{lj'} - \theta _{lkj'}$. On the other hand, similarly as (\ref{eq:internalStateDt}), we can write $dw_ {lkj'}/dt$ as
\begin{align}
\frac{dw_ {lkj'}}{dt} = \Big( \frac{d(|w _{lkj'}|)}{dt} + i |w _{lkj'}| \frac{d\theta _{lkj'}}{dt} \Big) e ^{i\theta _{lkj'}}
\end{align}
Hence,
\begin{align}\label{eq:weightChangeDevAngle}
\frac{1}{e ^{i\theta _{lkj'}}} \frac{dw_ {lkj'}}{dt} = \Big( \frac{d(|w _{lkj'}|)}{dt} + i |w _{lkj'}| \frac{d\theta _{lkj'}}{dt} \Big)
\end{align}
By (\ref{eq:weightChangeDevAngleAP}) and (\ref{eq:weightChangeDevAngle}), we derive a relationship, for general $j = j'$, expressed as
\begin{equation}\label{eq:weightFracRot}
\left[ 
\begin{array}{c} 
\frac{d(|w _{lkj}|)}{dt} \\
|w _{lkj}| \frac{d\theta _{lkj}}{dt}
\end{array} 
\right] 
=
\begin{bmatrix} 
\cos{ \theta ^{\rm rot} _{lkj}} & -\sin{ \theta ^{\rm rot} _{lkj}} \\ 
\sin{ \theta ^{\rm rot} _{lkj}} & \cos{ \theta ^{\rm rot} _{lkj}} 
\end{bmatrix} \left[ 
\begin{array}{c} 
\frac{dw ^{\rm a} _{lkj}}{dt} \\
\frac{dw ^{\rm p} _{lkj}}{dt}
\end{array}
\right]
\end{equation}
where
\begin{align}
    \theta ^{\rm rot} _{lkj} \equiv \theta _{(l+1)k} - \theta _{lj} - \theta _{lkj}
\end{align}
The explicit expression for the two change fractions can be written as
\begin{equation}
\left[ 
\begin{array}{c} 
\frac{dw ^{\rm a} _{lkj}}{dt} \\
\frac{dw ^{\rm p} _{lkj}}{dt}
\end{array} 
\right] 
=
\begin{bmatrix} 
\cos{\theta ^{\rm rot} _{lkj}} & \sin{ \theta ^{\rm rot} _{lkj}} \\ 
-\sin{ \theta ^{\rm rot} _{lkj}} & \cos{ \theta ^{\rm rot} _{lkj}} 
\end{bmatrix} \left[ 
\begin{array}{c} 
\frac{d(|w _{lkj}|)}{dt} \\
|w _{lkj}| \frac{d\theta _{lkj}}{dt}
\end{array}
\right]
\end{equation}
This reads the following rule of the weight changes:
\begin{align}\label{eq:descDirAmp}
\frac{d|w _{lkj}|}{dt} =& - \frac{\partial E _{(l+1)}}{\partial |w _{lkj}|} \nonumber \\
    =& -\Big( \frac{\partial E _{(l+1)}}{\partial w ^{\rm a} _{lkj}} \frac{\partial w ^{\rm a} _{lkj}}{\partial |w _{lkj}|} + \frac{\partial E _{(l+1)}}{\partial w ^{\rm p} _{lkj}} \frac{\partial w ^{\rm p} _{lkj}}{\partial |w _{lkj}|} \Big) \nonumber \\ 
	=&-\Big( \frac{\partial E _{(l+1)}}{\partial w ^{\rm a} _{lkj}} \cos{\theta ^{\rm rot} _{lkj}} - \frac{\partial E _{(l+1)}}{\partial w ^{\rm p} _{lkj}} \sin{\theta ^{\rm rot} _{lkj}} \Big)
\end{align}
\begin{align}\label{eq:descDirPha}
\frac{d\theta _{lkj}}{dt} =& - \frac{1}{|w _{lkj}|} \frac{\partial E _{(l+1)}}{\partial \theta _{lkj}} \nonumber \\
    =& -\frac{1}{|w _{lkj}|} \Big( \frac{\partial E _{(l+1)}}{\partial w ^{\rm a} _{lkj}} \frac{\partial w ^{\rm a} _{lkj}}{\partial \theta _{lkj}} + \frac{\partial E _{(l+1)}}{\partial w ^{\rm p} _{lkj}} \frac{\partial w ^{\rm p} _{lkj}}{\partial \theta _{lkj}} \Big) \nonumber \\ 
	=&-\Big( \frac{\partial E _{(l+1)}}{\partial w ^{\rm a} _{lkj}} \sin{\theta ^{\rm rot} _{lkj}} + \frac{\partial E _{(l+1)}}{\partial w ^{\rm p} _{lkj}} \cos{\theta ^{\rm rot} _{lkj}} \Big)
\end{align}

For the conventional objective function (\ref{eq:objectiveConv}) \citep{Hirose2012}, we get
\begin{align}
\frac{\partial E _{(l+1)}}{\partial w ^{\rm a} _{lkj}} =& \frac{\partial E  _{(l+1)}}{\partial (|u _{(l+1)k}|)}\frac{d(|u_ {(l+1)k}|)}{dw ^{\rm a} _{lkj}} \nonumber \\
=& (1 - |z _{(l+1)k}| ^2) \nonumber \\
& \times  \Big(|z _{(l+1)k}|-|\hat {z} _{(l+1)k}| \cos{(\theta _{(l+1)k} - \hat {\theta} _{(l+1)k})}\Big) |z _{lj}|
\end{align}
\begin{align}
\frac{\partial E _{(l+1)}}{\partial w ^{\rm p} _{lkj}} &= \frac{\partial E  _{(l+1)}}{\partial \theta _{(l+1)k}}\frac{d \theta _ {(l+1)k}}{dw ^{\rm p} _{lkj}} \nonumber \\
&= |z _{(l+1)k}||\hat {z} _{(l+1)k}| \sin{(\theta _{(l+1)k} - \hat {\theta} _{(l+1)k})} \frac{|z _{lj}|}{|u _{(l+1)k}|} 
\end{align}
since we have the relationships (\ref{eq:dabsudwa}) and (\ref{eq:dangleudwp}) as well as
\begin{align}
E _{(l+1)} \equiv \frac{1}{2}| & \mbox{\boldmath $z$} _{(l+1)} - \hat {\mbox{\boldmath $z$}} _{(l+1)} |^{2} \nonumber \\
=\frac{1}{2} & \sum _{k} | z _{(l+1)k} - \hat{z} _{(l+1)k} |^{2} \nonumber \\
=\frac{1}{2} & \sum _{k} \Big( {\rm tanh} ^{2} (|u _{(l+1)k}|) + {\rm tanh} ^{2}(|\hat {u} _{(l+1)k}|) \nonumber \\
& -2 {\rm tanh}(|u _{(l+1)k}|) {\rm tanh}(| \hat {u} _{(l+1)k}|) \cos{(\theta _{(l+1)k} - \hat {\theta} _{(l+1)k})} \Big)
\end{align}
where $\hat {\mbox{\boldmath $z$}} _{(l+1)} = [\hat{z} _{(l+1)k}]$ are the teacher signals of the outputs $\mbox{\boldmath $z$} _{(l+1)} = [z _{(l+1)k}]$ and $\hat {\mbox{\boldmath $u$}} _{(l+1)} = [|\hat{u} _{(l+1)k}|e ^{i \hat{\theta} _{(l+1)k}}]$ are the equivalent internal state corresponding to the teacher signals.

The objective function with the $L _{1}$-norm penalty is
\begin{align}
E ^{\rm S} _{(l+1)} &= \frac{1}{2} | \mbox{\boldmath $ z$}_{(l+1)} - \hat{\mbox{\boldmath $ z$}}_{(l+1)}|^2 + \alpha \| \mbox{\boldmath $ W $}_{l} \|_{1} \nonumber \\
&= E _{(l+1)} + \alpha \sum_k \sum_j |w _{lkj}|
\end{align}
where $\alpha$ is a penalty coefficient. Since $\partial (|w _{lkj}|)/ \partial w ^{\rm a} _{lkj} = \cos{\theta ^{\rm rot} _{lkj}}$ and $\partial (|w _{lkj}|)/ \partial w ^{\rm p} _{lkj} = - \sin{\theta ^{\rm rot} _{lkj}}$ by (\ref{eq:weightFracRot}), the descent direction of the steepest descent with $E ^{\rm S} _{(l+1)}$ can be written as
\begin{align}
\frac{d |w _{lkj}|}{dt} =& - \frac{\partial E ^{\rm S} _{(l+1)}}{\partial |w _{lkj}|} = -\Big( \frac{\partial E ^{\rm S} _{(l+1)}}{\partial w ^{\rm a} _{lkj}} \cos{\theta ^{\rm rot} _{lkj}} - \frac{\partial E ^{\rm S} _{(l+1)}}{\partial w ^{\rm p} _{lkj}} \sin{\theta ^{\rm rot} _{lkj}} \Big) \nonumber \\
= - \Bigg\{ \Big(& \frac{\partial E _{(l+1)}}{\partial w ^{\rm a} _{lkj}} + \alpha \frac{\partial |w _{lkj}|}{\partial w ^{\rm a} _{lkj}} \Big) \cos{\theta ^{\rm rot} _{lkj}}  - \Big(\frac{\partial E _{(l+1)}}{\partial w ^{\rm p} _{lkj}} + \alpha \frac{\partial |w _{lkj}|}{\partial w ^{\rm p} _{lkj}}  \Big) \sin{\theta ^{\rm rot} _{lkj}} \Bigg\} \nonumber \\
= - \Bigg\{  (&1 - |z _{(l+1)k}| ^2) \Big(|z _{(l+1)k}|-|\hat {z} _{(l+1)k}| \cos{(\theta _{(l+1)k} - \hat {\theta} _{(l+1)k})}\Big) |z _{lj}| \cos{\theta ^{\rm rot} _{lkj}} \nonumber \\
&- |z _{(l+1)k}||\hat {z} _{(l+1)k}| \sin{(\theta _{(l+1)k} - \hat {\theta} _{(l+1)k})} \frac{|z _{lj}|}{|u _{(l+1)k}|} \sin{\theta ^{\rm rot} _{lkj}} \nonumber \\
& +\alpha \big( \cos^{2}{\theta ^{\rm rot} _{lkj}} + \sin^{2}{\theta ^{\rm rot} _{lkj}} \big) \Bigg\}
\end{align}
\begin{align}
\frac{d \theta _{lkj}}{dt} =& - \frac{1}{|w _{lkj}|} \frac{\partial E ^{\rm S} _{(l+1)}}{\partial \theta _{lkj}} = -\Big( \frac{\partial E ^{\rm S} _{(l+1)}}{\partial w ^{\rm a} _{lkj}} \sin{\theta ^{\rm rot} _{lkj}} + \frac{\partial E ^{\rm S} _{(l+1)}}{\partial w ^{\rm p} _{lkj}} \cos{\theta ^{\rm rot} _{lkj}} \Big) \nonumber \\
= - \Bigg\{ \Big(& \frac{\partial E _{(l+1)}}{\partial w ^{\rm a} _{lkj}} + \alpha \frac{\partial |w _{lkj}|}{\partial w ^{\rm a} _{lkj}} \Big) \sin{\theta ^{\rm rot} _{lkj}}  + \Big(\frac{\partial E _{(l+1)}}{\partial w ^{\rm p} _{lkj}} + \alpha \frac{\partial |w _{lkj}|}{\partial w ^{\rm p} _{lkj}}  \Big) \cos{\theta ^{\rm rot} _{lkj}} \Bigg\} \nonumber \\
= - \Bigg\{  (&1 - |z _{(l+1)k}| ^2) \Big(|z _{(l+1)k}|-|\hat {z} _{(l+1)k}| \cos{(\theta _{(l+1)k} - \hat {\theta} _{(l+1)k})}\Big) |z _{lj}| \sin{\theta ^{\rm rot} _{lkj}} \nonumber \\
&+ |z _{(l+1)k}||\hat {z} _{(l+1)k}| \sin{(\theta _{(l+1)k} - \hat {\theta} _{(l+1)k})} \frac{|z _{lj}|}{|u _{(l+1)k}|} \cos{\theta ^{\rm rot} _{lkj}} \nonumber \\
& +\alpha \big( \cos{\theta ^{\rm rot} _{lkj}}\sin{\theta ^{\rm rot} _{lkj}} - \cos{\theta ^{\rm rot} _{lkj}}\sin{\theta ^{\rm rot} _{lkj}} \big) \Bigg\}
\end{align}
Therefore, we obtain (\ref{eq:weightRenewAmpSparse}) and (\ref{eq:weightRenewPhaSparse}).

For the objective function with the $L _{2,1}$-norm penalty,
\begin{align}
E ^{\rm GS} _{(l+1)} &= \frac{1}{2} | \mbox{\boldmath $ z$}_{(l+1)} - \hat{\mbox{\boldmath $ z$}}_{(l+1)}|^2 + \alpha \| \mbox{\boldmath $ W $}_{l} \|_{2,1} \nonumber \\
&= E _{(l+1)} + \alpha \sum _{j} \Big( \sqrt{|\boldsymbol{w}_{lj}|} \sqrt{\sum _{k} |w _{lkj}|^2} \Big)
\end{align}
where $\boldsymbol{w}_{lj} = [w _{lkj}]$ and $|\boldsymbol{w}_{lj}|$ denotes the dimensionality of the vector $\boldsymbol{w}_{lj}$. Based on (\ref{eq:weightFracRot}) and $w _{lkj} = |w _{lkj}| e ^{i\theta _{lkj}}$, we can get the following relationship:
\begin{align}\label{eq:dSquareAbsWdA}
    \frac{\partial |w _{lkj}| ^2}{\partial w ^{\rm a} _{lkj}} =& \frac{\partial w _{lkj}}{\partial w ^{\rm a} _{lkj}} w ^{*} _{lkj} + w _{lkj} \frac{\partial w^{*} _{lkj}}{\partial w ^{\rm a} _{lkj}} \nonumber \\
    = &\big( \frac{\partial w _{lkj}}{\partial |w _{lkj}|} \frac{\partial |w _{lkj}|}{\partial w ^{\rm a} _{lkj}} + \frac{\partial w _{lkj}}{\partial \theta _{lkj}} \frac{\partial \theta _{lkj}}{\partial w ^{\rm a} _{lkj}} \big) w^{*} _{lkj} \nonumber \\
    & + w _{lkj} \big(  \frac{\partial w^{*} _{lkj}}{\partial |w _{lkj}|} \frac{\partial |w _{lkj}|}{\partial w ^{\rm a} _{lkj}} + \frac{\partial w^{*} _{lkj}}{\partial \theta _{lkj}} \frac{\partial \theta _{lkj}}{\partial w ^{\rm a} _{lkj}} \big) \nonumber \\
    = & 2 |w _{lkj}| \cos{\theta ^{\rm rot} _{lkj}}
\end{align}
\begin{align}\label{eq:dSquareAbsWdP}
    \frac{\partial |w _{lkj}| ^2}{\partial w ^{\rm p} _{lkj}} =& \frac{\partial w _{lkj}}{\partial w ^{\rm p} _{lkj}} w^{*} _{lkj} + w _{lkj} \frac{\partial w^{*} _{lkj}}{\partial w ^{\rm p} _{lkj}} \nonumber \\
    = &\big( \frac{\partial w _{lkj}}{\partial |w _{lkj}|} \frac{\partial |w _{lkj}|}{\partial w ^{\rm p} _{lkj}} + \frac{\partial w _{lkj}}{\partial \theta _{lkj}} \frac{\partial \theta _{lkj}}{\partial w ^{\rm p} _{lkj}} \big) w^{*} _{lkj} \nonumber \\
    & + w _{lkj} \big(  \frac{\partial w^{*} _{lkj}}{\partial |w _{lkj}|} \frac{\partial |w _{lkj}|}{\partial w ^{\rm p} _{lkj}} + \frac{\partial w^{*} _{lkj}}{\partial \theta _{lkj}} \frac{\partial \theta _{lkj}}{\partial w ^{\rm p} _{lkj}} \big) \nonumber \\
    = & - 2 |w _{lkj}| \sin{\theta ^{\rm rot} _{lkj}}
\end{align}
By (\ref{eq:dSquareAbsWdA}) and (\ref{eq:dSquareAbsWdP}), the descent direction of the steepest descent with $E ^{\rm GS} _{(l+1)}$ can be written as
\begin{align}
\frac{d |w _{lkj}|}{dt} =& - \frac{\partial E ^{\rm GS} _{(l+1)}}{\partial |w _{lkj}|} = -\Big( \frac{\partial E ^{\rm GS} _{(l+1)}}{\partial w ^{\rm a} _{lkj}} \cos{\theta ^{\rm rot} _{lkj}} - \frac{\partial E ^{\rm GS} _{(l+1)}}{\partial w ^{\rm p} _{lkj}} \sin{\theta ^{\rm rot} _{lkj}} \Big) \nonumber \\
= - \Bigg\{ \Big(& \frac{\partial E _{(l+1)}}{\partial w ^{\rm a} _{lkj}} + \alpha \frac{\sqrt{|\boldsymbol{w}_{lj}|}}{2\| \boldsymbol{w} _{lj} \| _{2}} \frac{\partial |w _{lkj}| ^{2}}{\partial w ^{\rm a} _{lkj}} \Big) \cos{\theta ^{\rm rot} _{lkj}}  \nonumber \\
&- \Big(\frac{\partial E _{(l+1)}}{\partial w ^{\rm p} _{lkj}} + \alpha \frac{\sqrt{|\boldsymbol{w}_{lj}|}}{2\| \boldsymbol{w} _{lj} \| _{2}} \frac{\partial |w _{lkj}| ^{2}}{\partial w ^{\rm p} _{lkj}}  \Big) \sin{\theta ^{\rm rot} _{lkj}} \Bigg\} \nonumber \\
= - \Bigg\{  (&1 - |z _{(l+1)k}| ^2) \Big(|z _{(l+1)k}|-|\hat {z} _{(l+1)k}| \cos{(\theta _{(l+1)k} - \hat {\theta} _{(l+1)k})}\Big) |z _{lj}| \cos{\theta ^{\rm rot} _{lkj}} \nonumber \\
&- |z _{(l+1)k}||\hat {z} _{(l+1)k}| \sin{(\theta _{(l+1)k} - \hat {\theta} _{(l+1)k})} \frac{|z _{lj}|}{|u _{(l+1)k}|} \sin{\theta ^{\rm rot} _{lkj}} \nonumber \\
& +\alpha \sqrt{|\boldsymbol{w}_{lj}|} \frac{|w _{lkj}|}{\| \boldsymbol{w} _{lj} \| _{2}} \big( \cos^{2}{\theta ^{\rm rot} _{lkj}} + \sin^{2}{\theta ^{\rm rot} _{lkj}}\big) \Bigg\} 
\end{align}
\begin{align}
\frac{d \theta _{lkj}}{dt} =& - \frac{1}{|w _{lkj}|} \frac{\partial E ^{\rm GS} _{(l+1)}}{\partial \theta _{lkj}} = -\Big( \frac{\partial E ^{\rm GS} _{(l+1)}}{\partial w ^{\rm a} _{lkj}} \sin{\theta ^{\rm rot} _{lkj}} + \frac{\partial E ^{\rm GS} _{(l+1)}}{\partial w ^{\rm p} _{lkj}} \cos{\theta ^{\rm rot} _{lkj}} \Big) \nonumber \\
= - \Bigg\{ \Big(& \frac{\partial E _{(l+1)}}{\partial w ^{\rm a} _{lkj}} + \alpha \frac{\sqrt{|\boldsymbol{w}_{lj}|}}{2\| \boldsymbol{w} _{lj} \| _{2}} \frac{\partial |w _{lkj}| ^{2}}{\partial w ^{\rm a} _{lkj}} \Big) \sin{\theta ^{\rm rot} _{lkj}}  \nonumber \\
&+ \Big(\frac{\partial E _{(l+1)}}{\partial w ^{\rm p} _{lkj}} + \alpha \frac{\sqrt{|\boldsymbol{w}_{lj}|}}{2\| \boldsymbol{w} _{lj} \| _{2}} \frac{\partial |w _{lkj}| ^{2}}{\partial w ^{\rm p} _{lkj}}  \Big) \cos{\theta ^{\rm rot} _{lkj}} \Bigg\} \nonumber \\
= - \Bigg\{  (&1 - |z _{(l+1)k}| ^2) \Big(|z _{(l+1)k}|-|\hat {z} _{(l+1)k}| \cos{(\theta _{(l+1)k} - \hat {\theta} _{(l+1)k})}\Big) |z _{lj}| \sin{\theta ^{\rm rot} _{lkj}} \nonumber \\
&+ |z _{(l+1)k}||\hat {z} _{(l+1)k}| \sin{(\theta _{(l+1)k} - \hat {\theta} _{(l+1)k})} \frac{|z _{lj}|}{|u _{(l+1)k}|} \cos{\theta ^{\rm rot} _{lkj}} \nonumber \\
& +\alpha \sqrt{|\boldsymbol{w}_{lj}|} \frac{|w _{lkj}|}{\| \boldsymbol{w} _{lj} \| _{2}} \big( \cos{\theta ^{\rm rot} _{lkj}}\sin{\theta ^{\rm rot} _{lkj}} - \cos{\theta ^{\rm rot} _{lkj}}\sin{\theta ^{\rm rot} _{lkj}} \big) \Bigg\} 
\end{align}
Therefore, we obtain (\ref{eq:weightRenewAmpGroupSparse}) and (\ref{eq:weightRenewPhaGroupSparse}).

Note that we, in this work, use the steepest descent on the regularization problems, which may only guarantee significantly small weights instead of providing exact-zero amplitude for redundant weight connections. Although significantly small versus exactly zero weights have limited difference on the performance of the channel prediction, that is prediction accuracy, one may want exact-zero weights or completely pruned neurons for reduction of power consumption in a practical mobile communications. A possible extension of the proposed methods by utilizing other proximal algorithms such as iterative shrinkage-thresholding algorithm (ISTA) in the optimization of the regularization \citep{Gregor2010,Li2015} is an interesting topic but we leave this to future investigation because this is out of the scope of this paper.  

\section{Structure of long short-term memory network}
\label{appendix:lstmNet}
As a performance comparision of channel prediction, a simple long short-term memory (LSTM) network \citep{Cho2014}, a variation of real-valued recurrent neural networks, is used in this work. The mathematical description of the LSTM network is as follows:
\begin{align}
    \mbox{\boldmath $y$}_{t} &= \sigma(\mbox{\boldmath $ w$}_{y} \mbox{\boldmath $ x$}_{t} + \mbox{\boldmath $U$}_{y}\mbox{\boldmath $ h$}_{(t-1)}) \\
    \mbox{\boldmath $r$}_{t} &= \sigma(\mbox{\boldmath $ w$}_{r} \mbox{\boldmath $ x$}_{t} + \mbox{\boldmath $U$}_{r}\mbox{\boldmath $ h$}_{(t-1)}) \\
    \tilde{\mbox{\boldmath $h$}}_{t} &= \tanh{\big(\mbox{\boldmath $ w$} \mbox{\boldmath $ x$}_{t} + \mbox{\boldmath $U$}(\mbox{\boldmath $ r$}_{t} \odot \mbox{\boldmath $ h$}_{(t-1)})\big)} \\
    \mbox{\boldmath $h$}_{t} &= \mbox{\boldmath $ y$}_{t} \odot \mbox{\boldmath $ h$}_{(t-1)} + (1-\mbox{\boldmath $y$}_{t}) \odot \tilde{\mbox{\boldmath $h$}}_{t}
\end{align}
where $\mbox{\boldmath $ x$}_{t}$, $\mbox{\boldmath $ h$}_{(t-1)}$, $\mbox{\boldmath $y$}_{t}$, $\mbox{\boldmath $r$}_{t}$, $\tilde{\mbox{\boldmath $h$}}_{t}$ and $\mbox{\boldmath $h$}_{t}$ are inputs, previous hidden states, update gates, reset gates, intermediate hidden states and present hidden states, respectively, $\mbox{\boldmath $ w$}_{y}$, $\mbox{\boldmath $U$}_{y}$, $\mbox{\boldmath $ w$}_{r}$, $\mbox{\boldmath $U$}_{r}$, $\mbox{\boldmath $ w$}$ and $\mbox{\boldmath $U$}$ are weight matrices which are learned, and $\odot$ denotes the Hadamard product of vectors and $\sigma(x) = 1/(1+e^{-x})$. A set of estimated past path characteristics $\hat{c} _{m} (t-1), ..., \hat{c} _{m} (t-I_{\rm ML})$ is converted into the input vector $\mbox{\boldmath $ x$}_{t}$ by splitting real and imaginary parts of the complex value. The weight matrices are updated by using the steepest descent method with the standard error-backpropagation so that they minimize the difference
\begin{equation}\label{eq:objectiveLSTM}
E ^{\rm lstm} \equiv \frac{1}{2}| \mbox{\boldmath $h$} _{t} - \hat {\mbox{\boldmath $h$}} _{t} |^{2}
\end{equation}
where $\hat {\mbox{\boldmath $h$}} _{t}$ denotes the teacher signals of $\mbox{\boldmath $ h$}_{t}$ consists of real and imaginary parts of $\hat{c} _{m} (t)$. The learned weights are kept internally in the network and updated in the following time points until the latest channel component is used. The future channel states are predicted by the most up-to-dated weights.    

\end{appendix}


\bibliography{Channel_prediction_using_CVNN.bib}

\begin{thebibliography}{54}
\expandafter\ifx\csname natexlab\endcsname\relax\def\natexlab#1{#1}\fi
\providecommand{\url}[1]{\texttt{#1}}
\providecommand{\href}[2]{#2}
\providecommand{\path}[1]{#1}
\providecommand{\DOIprefix}{doi:}
\providecommand{\ArXivprefix}{arXiv:}
\providecommand{\URLprefix}{URL: }
\providecommand{\Pubmedprefix}{pmid:}
\providecommand{\doi}[1]{\href{http://dx.doi.org/#1}{\path{#1}}}
\providecommand{\Pubmed}[1]{\href{pmid:#1}{\path{#1}}}
\providecommand{\bibinfo}[2]{#2}
\ifx\xfnm\relax \def\xfnm[#1]{\unskip,\space#1}\fi
\bibitem[{Aghasi et~al.(2017)Aghasi, Abdi, Nguyen \& Romberg}]{Aghasi2017}
\bibinfo{author}{Aghasi, A.}, \bibinfo{author}{Abdi, A.},
  \bibinfo{author}{Nguyen, N.}, \& \bibinfo{author}{Romberg, J.}
  (\bibinfo{year}{2017}).
\newblock \bibinfo{title}{{Net-Trim: Convex pruning of deep neural networks
  with performance guarantee}}.
\newblock {\it \bibinfo{journal}{Advances in Neural Information Processing
  Systems}\/},  {\it \bibinfo{volume}{2017-Decem}\/},
  \bibinfo{pages}{3178--3187}.
\bibitem[{Arima \& Hirose(2017)}]{Arima2017}
\bibinfo{author}{Arima, Y.}, \& \bibinfo{author}{Hirose, A.}
  (\bibinfo{year}{2017}).
\newblock \bibinfo{title}{{Performance Dependence on System Parameters in
  Millimeter-Wave Active Imaging Based on Complex-Valued Neural Networks to
  Classify Complex Texture}}.
\newblock {\it \bibinfo{journal}{IEEE Access}\/},  {\it \bibinfo{volume}{5}\/},
  \bibinfo{pages}{22927--22939}. \DOIprefix\doi{10.1109/ACCESS.2017.2751618}.
\bibitem[{Arredondo et~al.(2002)Arredondo, Dandekar \& {Guanghan
  Xu}}]{Arredondo2002}
\bibinfo{author}{Arredondo, A.}, \bibinfo{author}{Dandekar, K.}, \&
  \bibinfo{author}{{Guanghan Xu}} (\bibinfo{year}{2002}).
\newblock \bibinfo{title}{{Vector channel modeling and prediction for the
  improvement of downlink received power}}.
\newblock {\it \bibinfo{journal}{IEEE Transactions on Communications}\/},  {\it
  \bibinfo{volume}{50}\/}, \bibinfo{pages}{1121--1129}.
  \DOIprefix\doi{10.1109/TCOMM.2002.800827}.
\bibitem[{Barakat et~al.(2011)Barakat, Druaux, Lefebvre, Khalil \&
  Mustapha}]{Barakat2011}
\bibinfo{author}{Barakat, M.}, \bibinfo{author}{Druaux, F.},
  \bibinfo{author}{Lefebvre, D.}, \bibinfo{author}{Khalil, M.}, \&
  \bibinfo{author}{Mustapha, O.} (\bibinfo{year}{2011}).
\newblock \bibinfo{title}{{Self adaptive growing neural network classifier for
  faults detection and diagnosis}}.
\newblock {\it \bibinfo{journal}{Neurocomputing}\/},  {\it
  \bibinfo{volume}{74}\/}, \bibinfo{pages}{3865--3876}.
  \DOIprefix\doi{10.1016/j.neucom.2011.08.001}.
\bibitem[{Bui et~al.(2013)Bui, Ogawa, Nishimura \& Ohgane}]{Bui2013}
\bibinfo{author}{Bui, H.~P.}, \bibinfo{author}{Ogawa, Y.},
  \bibinfo{author}{Nishimura, T.}, \& \bibinfo{author}{Ohgane, T.}
  (\bibinfo{year}{2013}).
\newblock \bibinfo{title}{{Performance Evaluation of a Multi-User MIMO System
  With Prediction of Time-Varying Indoor Channels}}.
\newblock {\it \bibinfo{journal}{IEEE Transactions on Antennas and
  Propagation}\/},  {\it \bibinfo{volume}{61}\/}, \bibinfo{pages}{371--379}.
  \DOIprefix\doi{10.1109/TAP.2012.2214995}.
\bibitem[{Bui et~al.(2017)Bui, Cesana, Hosseini, Liao, Malanchini \&
  Widmer}]{Bui2017}
\bibinfo{author}{Bui, N.}, \bibinfo{author}{Cesana, M.},
  \bibinfo{author}{Hosseini, S.~A.}, \bibinfo{author}{Liao, Q.},
  \bibinfo{author}{Malanchini, I.}, \& \bibinfo{author}{Widmer, J.}
  (\bibinfo{year}{2017}).
\newblock \bibinfo{title}{{A Survey of Anticipatory Mobile Networking:
  Context-Based Classification, Prediction Methodologies, and Optimization
  Techniques}}.
\newblock {\it \bibinfo{journal}{IEEE Communications Surveys {\&}
  Tutorials}\/},  {\it \bibinfo{volume}{19}\/}, \bibinfo{pages}{1790--1821}.
  \DOIprefix\doi{10.1109/COMST.2017.2694140}.
  \href{http://arxiv.org/abs/1606.00191}{\tt arXiv:1606.00191}.
\bibitem[{Cand{\`{e}}s et~al.(2006)Cand{\`{e}}s, Romberg \& Tao}]{Candes2006}
\bibinfo{author}{Cand{\`{e}}s, E.~J.}, \bibinfo{author}{Romberg, J.~K.}, \&
  \bibinfo{author}{Tao, T.} (\bibinfo{year}{2006}).
\newblock \bibinfo{title}{{Stable signal recovery from incomplete and
  inaccurate measurements}}.
\newblock {\it \bibinfo{journal}{Communications on Pure and Applied
  Mathematics}\/},  {\it \bibinfo{volume}{59}\/}, \bibinfo{pages}{1207--1223}.
  \DOIprefix\doi{10.1002/cpa.20124}. \href{http://arxiv.org/abs/0503066}{\tt
  arXiv:0503066}.
\bibitem[{Cho et~al.(2014)Cho, van Merrienboer, Gulcehre, Bahdanau, Bougares,
  Schwenk \& Bengio}]{Cho2014}
\bibinfo{author}{Cho, K.}, \bibinfo{author}{van Merrienboer, B.},
  \bibinfo{author}{Gulcehre, C.}, \bibinfo{author}{Bahdanau, D.},
  \bibinfo{author}{Bougares, F.}, \bibinfo{author}{Schwenk, H.}, \&
  \bibinfo{author}{Bengio, Y.} (\bibinfo{year}{2014}).
\newblock \bibinfo{title}{{Learning Phrase Representations using RNN
  Encoder–Decoder for Statistical Machine Translation}}.
\newblock In {\it \bibinfo{booktitle}{Proceedings of the 2014 Conference on
  Empirical Methods in Natural Language Processing (EMNLP)}\/} (pp.
  \bibinfo{pages}{1724--1734}).
\newblock \bibinfo{address}{Stroudsburg, PA, USA}:
  \bibinfo{publisher}{Association for Computational Linguistics}
\newblock volume~\bibinfo{volume}{28}. \DOIprefix\doi{10.3115/v1/D14-1179}.
\bibitem[{Cho et~al.(2010)Cho, Kim, Yang \& Kang}]{Cho2010}
\bibinfo{author}{Cho, Y.~S.}, \bibinfo{author}{Kim, J.}, \bibinfo{author}{Yang,
  W.~Y.}, \& \bibinfo{author}{Kang, C.~G.} (\bibinfo{year}{2010}).
\newblock {\it \bibinfo{title}{{MIMO-OFDM Wireless Communications with
  MATLAB{\textregistered}}}\/}.
\newblock
\newblock \bibinfo{address}{Chichester, UK}: \bibinfo{publisher}{John Wiley
  {\&} Sons, Ltd}. \DOIprefix\doi{10.1002/9780470825631}.
  \href{http://arxiv.org/abs/arXiv:1011.1669v3}{\tt arXiv:arXiv:1011.1669v3}.
\bibitem[{Ding \& Hirose(2014{\natexlab{a}})}]{Ding2014b}
\bibinfo{author}{Ding, T.}, \& \bibinfo{author}{Hirose, A.}
  (\bibinfo{year}{2014}{\natexlab{a}}).
\newblock \bibinfo{title}{{Fading Channel Prediction Based on Combination of
  Complex-Valued Neural Networks and Chirp Z-Transform}}.
\newblock {\it \bibinfo{journal}{IEEE Transactions on Neural Networks and
  Learning Systems}\/},  {\it \bibinfo{volume}{25}\/},
  \bibinfo{pages}{1686--1695}. \DOIprefix\doi{10.1109/TNNLS.2014.2306420}.
\bibitem[{Ding \& Hirose(2014{\natexlab{b}})}]{Ding2014a}
\bibinfo{author}{Ding, T.}, \& \bibinfo{author}{Hirose, A.}
  (\bibinfo{year}{2014}{\natexlab{b}}).
\newblock \bibinfo{title}{{Fading Channel Prediction Based on Self-optimizing
  Neural Networks}}.
\newblock In {\it \bibinfo{booktitle}{Lecture Notes in Computer Science}\/}
  (pp. \bibinfo{pages}{175--182}).
\newblock
\newblock volume \bibinfo{volume}{8834}.
  \DOIprefix\doi{10.1007/978-3-319-12637-1_22}.
\bibitem[{Donoho \& Elad(2003)}]{Donoho2003}
\bibinfo{author}{Donoho, D.~L.}, \& \bibinfo{author}{Elad, M.}
  (\bibinfo{year}{2003}).
\newblock \bibinfo{title}{{Optimally sparse representation in general
  (nonorthogonal) dictionaries via 1 minimization}}.
\newblock {\it \bibinfo{journal}{Proceedings of the National Academy of
  Sciences}\/},  {\it \bibinfo{volume}{100}\/}, \bibinfo{pages}{2197--2202}.
  \DOIprefix\doi{10.1073/pnas.0437847100}.
\bibitem[{Donoho \& Tanner(2008)}]{Donoho2008}
\bibinfo{author}{Donoho, D.~L.}, \& \bibinfo{author}{Tanner, J.}
  (\bibinfo{year}{2008}).
\newblock \bibinfo{title}{{Counting faces of randomly projected polytopes when
  the projection radically lowers dimension}}.
\newblock {\it \bibinfo{journal}{Journal of the American Mathematical
  Society}\/},  {\it \bibinfo{volume}{22}\/}, \bibinfo{pages}{1--53}.
  \DOIprefix\doi{10.1090/S0894-0347-08-00600-0}.
\bibitem[{Duel-Hallen(2007)}]{Duel-Hallen2007}
\bibinfo{author}{Duel-Hallen, A.} (\bibinfo{year}{2007}).
\newblock \bibinfo{title}{{Fading Channel Prediction for Mobile Radio Adaptive
  Transmission Systems}}.
\newblock {\it \bibinfo{journal}{Proceedings of the IEEE}\/},  {\it
  \bibinfo{volume}{95}\/}, \bibinfo{pages}{2299--2313}.
  \DOIprefix\doi{10.1109/JPROC.2007.904443}.
\bibitem[{Duel-Hallen et~al.(2006)Duel-Hallen, Hallen \& {Tung-Sheng
  Yang}}]{Duel-Hallen2006}
\bibinfo{author}{Duel-Hallen, A.}, \bibinfo{author}{Hallen, H.}, \&
  \bibinfo{author}{{Tung-Sheng Yang}} (\bibinfo{year}{2006}).
\newblock \bibinfo{title}{{Long range prediction and reduced feedback for
  mobile radio adaptive OFDM systems}}.
\newblock {\it \bibinfo{journal}{IEEE Transactions on Wireless
  Communications}\/},  {\it \bibinfo{volume}{5}\/},
  \bibinfo{pages}{2723--2733}. \DOIprefix\doi{10.1109/TWC.2006.04219}.
\bibitem[{Elad(2010)}]{Elad2010}
\bibinfo{author}{Elad, M.} (\bibinfo{year}{2010}).
\newblock {\it \bibinfo{title}{{Sparse and redundant representations: From
  theory to applications in signal and image processing}}\/}.
\newblock
\newblock \bibinfo{publisher}{Springer}.
  \DOIprefix\doi{10.1007/978-1-4419-7011-4}. \href{http://arxiv.org/abs/g}{\tt
  arXiv:g}.
\bibitem[{Elman(1993)}]{Elman1993}
\bibinfo{author}{Elman, J.~L.} (\bibinfo{year}{1993}).
\newblock \bibinfo{title}{{Learning and development in neural networks: the
  importance of starting small}}.
\newblock {\it \bibinfo{journal}{Cognition}\/},  {\it \bibinfo{volume}{48}\/},
  \bibinfo{pages}{71--99}. \DOIprefix\doi{10.1016/0010-0277(93)90058-4}.
\bibitem[{Eraslan et~al.(2013)Eraslan, Daneshrad \& Lou}]{Eraslan2013}
\bibinfo{author}{Eraslan, E.}, \bibinfo{author}{Daneshrad, B.}, \&
  \bibinfo{author}{Lou, C.-Y.} (\bibinfo{year}{2013}).
\newblock \bibinfo{title}{{Performance Indicator for MIMO MMSE Receivers in the
  Presence of Channel Estimation Error}}.
\newblock {\it \bibinfo{journal}{IEEE Wireless Communications Letters}\/},
  {\it \bibinfo{volume}{2}\/}, \bibinfo{pages}{211--214}.
  \DOIprefix\doi{10.1109/WCL.2013.012513.120824}.
  \href{http://arxiv.org/abs/1210.8191}{\tt arXiv:1210.8191}.
\bibitem[{Eyceoz et~al.(1998)Eyceoz, Duel-Hallen \& Hallen}]{Eyceoz1998}
\bibinfo{author}{Eyceoz, T.}, \bibinfo{author}{Duel-Hallen, A.}, \&
  \bibinfo{author}{Hallen, H.} (\bibinfo{year}{1998}).
\newblock \bibinfo{title}{{Deterministic channel modeling and long range
  prediction of fast fading mobile radio channels}}.
\newblock {\it \bibinfo{journal}{IEEE Communications Letters}\/},  {\it
  \bibinfo{volume}{2}\/}, \bibinfo{pages}{254--256}.
  \DOIprefix\doi{10.1109/4234.718494}.
\bibitem[{Gregor \& LeCun(2010)}]{Gregor2010}
\bibinfo{author}{Gregor, K.}, \& \bibinfo{author}{LeCun, Y.}
  (\bibinfo{year}{2010}).
\newblock \bibinfo{title}{{Learning fast approximations of sparse coding}}.
\newblock In {\it \bibinfo{booktitle}{Proceedings, 27th International
  Conference on Machine Learning}\/}
\newblock (pp. \bibinfo{pages}{399--406}).
\bibitem[{Gribonval \& Nielsen(2003)}]{Gribonval2003}
\bibinfo{author}{Gribonval, R.}, \& \bibinfo{author}{Nielsen, M.}
  (\bibinfo{year}{2003}).
\newblock \bibinfo{title}{{Sparse representations in unions of bases}}.
\newblock {\it \bibinfo{journal}{IEEE Transactions on Information Theory}\/},
  {\it \bibinfo{volume}{49}\/}, \bibinfo{pages}{3320--3325}.
  \DOIprefix\doi{10.1109/TIT.2003.820031}.
\bibitem[{Hara \& Hirose(2004)}]{Hara2004}
\bibinfo{author}{Hara, T.}, \& \bibinfo{author}{Hirose, A.}
  (\bibinfo{year}{2004}).
\newblock \bibinfo{title}{{Plastic mine detecting radar system using
  complex-valued self-organizing map that deals with multiple-frequency
  interferometric images}}.
\newblock {\it \bibinfo{journal}{Neural Networks}\/},  {\it
  \bibinfo{volume}{17}\/}, \bibinfo{pages}{1201--1210}.
  \DOIprefix\doi{10.1016/j.neunet.2004.07.012}.
\bibitem[{Hirose(1994)}]{Hirose1994}
\bibinfo{author}{Hirose, A.} (\bibinfo{year}{1994}).
\newblock \bibinfo{title}{{Applications of complex-valued neural networks to
  coherent optical computing using phase-sensitive detection scheme}}.
\newblock {\it \bibinfo{journal}{Information Sciences - Applications}\/},  {\it
  \bibinfo{volume}{2}\/}, \bibinfo{pages}{103--117}.
  \DOIprefix\doi{10.1016/1069-0115(94)90014-0}.
\bibitem[{Hirose(2012)}]{Hirose2012}
\bibinfo{author}{Hirose, A.} (\bibinfo{year}{2012}).
\newblock {\it \bibinfo{title}{{Complex-valued neural networks}}\/} volume
  \bibinfo{volume}{400}.
\newblock (\bibinfo{edition}{2nd} ed.).
\newblock
\newblock \bibinfo{address}{New York}: \bibinfo{publisher}{Springer-Verlag}.
  \DOIprefix\doi{10.1007/978-3-642-27632-3}.
\bibitem[{Hirose \& Eckmiller(1996)}]{Hirose1996}
\bibinfo{author}{Hirose, A.}, \& \bibinfo{author}{Eckmiller, R.}
  (\bibinfo{year}{1996}).
\newblock \bibinfo{title}{{Coherent optical neural networks that have
  optical-frequency-controlled behavior and generalization ability in the
  frequency domain}}.
\newblock {\it \bibinfo{journal}{Applied Optics}\/},  {\it
  \bibinfo{volume}{35}\/}, \bibinfo{pages}{836}.
  \DOIprefix\doi{10.1364/AO.35.000836}.
\bibitem[{Hirose \& Yoshida(2012)}]{Hirose2012a}
\bibinfo{author}{Hirose, A.}, \& \bibinfo{author}{Yoshida, S.}
  (\bibinfo{year}{2012}).
\newblock \bibinfo{title}{{Generalization Characteristics of Complex-Valued
  Feedforward Neural Networks in Relation to Signal Coherence}}.
\newblock {\it \bibinfo{journal}{IEEE Transactions on Neural Networks and
  Learning Systems}\/},  {\it \bibinfo{volume}{23}\/},
  \bibinfo{pages}{541--551}. \DOIprefix\doi{10.1109/TNNLS.2012.2183613}.
\bibitem[{Ho et~al.(2017)Ho, Ngo, Matthaiou \& Duong}]{Ho2017}
\bibinfo{author}{Ho, C.~D.}, \bibinfo{author}{Ngo, H.~Q.},
  \bibinfo{author}{Matthaiou, M.}, \& \bibinfo{author}{Duong, T.~Q.}
  (\bibinfo{year}{2017}).
\newblock \bibinfo{title}{{On the Performance of Zero-Forcing Processing in
  Multi-Way Massive MIMO Relay Networks}}.
\newblock {\it \bibinfo{journal}{IEEE Communications Letters}\/},  {\it
  \bibinfo{volume}{21}\/}, \bibinfo{pages}{849--852}.
  \DOIprefix\doi{10.1109/LCOMM.2017.2648795}.
  \href{http://arxiv.org/abs/1701.00645}{\tt arXiv:1701.00645}.
\bibitem[{Hochreiter \& Schmidhuber(1997)}]{Hochreiter1997}
\bibinfo{author}{Hochreiter, S.}, \& \bibinfo{author}{Schmidhuber, J.}
  (\bibinfo{year}{1997}).
\newblock \bibinfo{title}{{Long Short-Term Memory}}.
\newblock {\it \bibinfo{journal}{Neural Computation}\/},  {\it
  \bibinfo{volume}{9}\/}, \bibinfo{pages}{1735--1780}.
  \DOIprefix\doi{10.1162/neco.1997.9.8.1735}.
  \href{http://arxiv.org/abs/1206.2944}{\tt arXiv:1206.2944}.
\bibitem[{Ishikawa(1996)}]{Ishikawa1996}
\bibinfo{author}{Ishikawa, M.} (\bibinfo{year}{1996}).
\newblock \bibinfo{title}{{Structural learning with forgetting}}.
\newblock {\it \bibinfo{journal}{Neural Networks}\/},  {\it
  \bibinfo{volume}{9}\/}, \bibinfo{pages}{509--521}.
  \DOIprefix\doi{10.1016/0893-6080(96)83696-3}.
\bibitem[{Jakes(1994)}]{Jakes1994}
\bibinfo{author}{Jakes, W.} (\bibinfo{year}{1994}).
\newblock {\it \bibinfo{title}{{Microwave Mobile Communications}}\/}.
\newblock (\bibinfo{edition}{2nd} ed.).
\newblock
\newblock \bibinfo{address}{New York}: \bibinfo{publisher}{Wiley}.
  \DOIprefix\doi{10.1109/9780470545287.ch1}.
\bibitem[{Karnin(1990)}]{Karnin1990}
\bibinfo{author}{Karnin, E.} (\bibinfo{year}{1990}).
\newblock \bibinfo{title}{{A simple procedure for pruning back-propagation
  trained neural networks}}.
\newblock {\it \bibinfo{journal}{IEEE Transactions on Neural Networks}\/},
  {\it \bibinfo{volume}{1}\/}, \bibinfo{pages}{239--242}.
  \DOIprefix\doi{10.1109/72.80236}.
\bibitem[{Kawata \& Hirose(2005)}]{Kawata2005}
\bibinfo{author}{Kawata, S.}, \& \bibinfo{author}{Hirose, A.}
  (\bibinfo{year}{2005}).
\newblock \bibinfo{title}{{Frequency-multiplexed logic circuit based on a
  coherent optical neural network}}.
\newblock {\it \bibinfo{journal}{Applied Optics}\/},  {\it
  \bibinfo{volume}{44}\/}, \bibinfo{pages}{4053}.
  \DOIprefix\doi{10.1364/AO.44.004053}.
\bibitem[{Koneru \& Vasudevan(2019)}]{Koneru2019}
\bibinfo{author}{Koneru, B. N.~G.}, \& \bibinfo{author}{Vasudevan, V.}
  (\bibinfo{year}{2019}).
\newblock \bibinfo{title}{{Sparse Artificial Neural Networks Using a Novel
  Smoothed LASSO Penalization}}.
\newblock {\it \bibinfo{journal}{IEEE Transactions on Circuits and Systems II:
  Express Briefs}\/},  {\it \bibinfo{volume}{66}\/}, \bibinfo{pages}{848--852}.
  \DOIprefix\doi{10.1109/TCSII.2019.2908729}.
\bibitem[{Li et~al.(2015)Li, Ding \& Li}]{Li2015}
\bibinfo{author}{Li, Z.}, \bibinfo{author}{Ding, S.}, \& \bibinfo{author}{Li,
  Y.} (\bibinfo{year}{2015}).
\newblock \bibinfo{title}{{A Fast Algorithm for Learning Overcomplete
  Dictionary for Sparse Representation Based on Proximal Operators}}.
\newblock {\it \bibinfo{journal}{Neural Computation}\/},  {\it
  \bibinfo{volume}{27}\/}, \bibinfo{pages}{1951--1982}.
  \DOIprefix\doi{10.1162/NECO_a_00763}.
  \href{http://arxiv.org/abs/1803.01446}{\tt arXiv:1803.01446}.
\bibitem[{Liu et~al.(2006)Liu, Yang \& Lajos}]{Liu2006}
\bibinfo{author}{Liu, W.}, \bibinfo{author}{Yang, L.-L.}, \&
  \bibinfo{author}{Lajos, H.} (\bibinfo{year}{2006}).
\newblock \bibinfo{title}{{Recurrent Neural Network Based Narrowband Channel
  Prediction}}.
\newblock In {\it \bibinfo{booktitle}{2006 IEEE 63rd Vehicular Technology
  Conference}\/} (pp. \bibinfo{pages}{2173--2177}).
\newblock \bibinfo{publisher}{IEEE}
\newblock volume~\bibinfo{volume}{5}.
  \DOIprefix\doi{10.1109/VETECS.2006.1683241}.
\bibitem[{Luo et~al.(2018)Luo, Ji, Wang, Chen \& Li}]{Luo2018}
\bibinfo{author}{Luo, C.}, \bibinfo{author}{Ji, J.}, \bibinfo{author}{Wang,
  Q.}, \bibinfo{author}{Chen, X.}, \& \bibinfo{author}{Li, P.}
  (\bibinfo{year}{2018}).
\newblock \bibinfo{title}{{Channel State Information Prediction for 5G Wireless
  Communications: A Deep Learning Approach}}.
\newblock {\it \bibinfo{journal}{IEEE Transactions on Network Science and
  Engineering}\/},  {\it \bibinfo{volume}{PP}\/}, \bibinfo{pages}{1--1}.
  \DOIprefix\doi{10.1109/TNSE.2018.2848960}.
\bibitem[{Maehara et~al.(2003)Maehara, Sasamori \& Tkahata}]{Maehara2003}
\bibinfo{author}{Maehara, F.}, \bibinfo{author}{Sasamori, F.}, \&
  \bibinfo{author}{Tkahata, F.} (\bibinfo{year}{2003}).
\newblock \bibinfo{title}{{Linear predictive maximal ratio combining
  transmitter diversity for OFDM-TDMA/TDD systems}}.
\newblock {\it \bibinfo{journal}{IEICE Transactions on Communications}\/},
  {\it \bibinfo{volume}{E86-B}\/}, \bibinfo{pages}{221--229}.
\bibitem[{O'Shea \& Hoydis(2017)}]{OShea2017}
\bibinfo{author}{O'Shea, T.}, \& \bibinfo{author}{Hoydis, J.}
  (\bibinfo{year}{2017}).
\newblock \bibinfo{title}{{An Introduction to Deep Learning for the Physical
  Layer}}.
\newblock {\it \bibinfo{journal}{IEEE Transactions on Cognitive Communications
  and Networking}\/},  {\it \bibinfo{volume}{3}\/}, \bibinfo{pages}{563--575}.
  \DOIprefix\doi{10.1109/TCCN.2017.2758370}.
  \href{http://arxiv.org/abs/1702.00832}{\tt arXiv:1702.00832}.
\bibitem[{Potter et~al.(2010)Potter, Venayagamoorthy \& Kosbar}]{Potter2010}
\bibinfo{author}{Potter, C.}, \bibinfo{author}{Venayagamoorthy, G.~K.}, \&
  \bibinfo{author}{Kosbar, K.} (\bibinfo{year}{2010}).
\newblock \bibinfo{title}{{RNN based MIMO channel prediction}}.
\newblock {\it \bibinfo{journal}{Signal Processing}\/},  {\it
  \bibinfo{volume}{90}\/}, \bibinfo{pages}{440--450}.
  \DOIprefix\doi{10.1016/j.sigpro.2009.07.013}.
\bibitem[{Ramachandram \& Taylor(2017)}]{Ramachandram2017}
\bibinfo{author}{Ramachandram, D.}, \& \bibinfo{author}{Taylor, G.~W.}
  (\bibinfo{year}{2017}).
\newblock \bibinfo{title}{{Deep Multimodal Learning: A Survey on Recent
  Advances and Trends}}.
\newblock {\it \bibinfo{journal}{IEEE Signal Processing Magazine}\/},  {\it
  \bibinfo{volume}{34}\/}, \bibinfo{pages}{96--108}.
  \DOIprefix\doi{10.1109/MSP.2017.2738401}.
\bibitem[{Reed(1993)}]{Reed1993}
\bibinfo{author}{Reed, R.} (\bibinfo{year}{1993}).
\newblock \bibinfo{title}{{Pruning algorithms-a survey}}.
\newblock {\it \bibinfo{journal}{IEEE Transactions on Neural Networks}\/},
  {\it \bibinfo{volume}{4}\/}, \bibinfo{pages}{740--747}.
  \DOIprefix\doi{10.1109/72.248452}.
\bibitem[{Ren et~al.(2018)Ren, Wu, Johansson, Shi \& Shi}]{Ren2018}
\bibinfo{author}{Ren, X.}, \bibinfo{author}{Wu, J.},
  \bibinfo{author}{Johansson, K.~H.}, \bibinfo{author}{Shi, G.}, \&
  \bibinfo{author}{Shi, L.} (\bibinfo{year}{2018}).
\newblock \bibinfo{title}{{Infinite Horizon Optimal Transmission Power Control
  for Remote State Estimation Over Fading Channels}}.
\newblock {\it \bibinfo{journal}{IEEE Transactions on Automatic Control}\/},
  {\it \bibinfo{volume}{63}\/}, \bibinfo{pages}{85--100}.
  \DOIprefix\doi{10.1109/TAC.2017.2709914}.
  \href{http://arxiv.org/abs/1604.08680}{\tt arXiv:1604.08680}.
\bibitem[{Scardapane et~al.(2017)Scardapane, Comminiello, Hussain \&
  Uncini}]{Scardapane2017}
\bibinfo{author}{Scardapane, S.}, \bibinfo{author}{Comminiello, D.},
  \bibinfo{author}{Hussain, A.}, \& \bibinfo{author}{Uncini, A.}
  (\bibinfo{year}{2017}).
\newblock \bibinfo{title}{{Group sparse regularization for deep neural
  networks}}.
\newblock {\it \bibinfo{journal}{Neurocomputing}\/},  {\it
  \bibinfo{volume}{241}\/}, \bibinfo{pages}{81--89}.
  \DOIprefix\doi{10.1016/j.neucom.2017.02.029}.
  \href{http://arxiv.org/abs/1607.00485}{\tt arXiv:1607.00485}.
\bibitem[{Sharma \& Chandra(2007)}]{Sharma2007}
\bibinfo{author}{Sharma, P.}, \& \bibinfo{author}{Chandra, K.}
  (\bibinfo{year}{2007}).
\newblock \bibinfo{title}{{Prediction of State Transitions in Rayleigh Fading
  Channels}}.
\newblock {\it \bibinfo{journal}{IEEE Transactions on Vehicular Technology}\/},
   {\it \bibinfo{volume}{56}\/}, \bibinfo{pages}{416--425}.
  \DOIprefix\doi{10.1109/TVT.2007.891421}.
\bibitem[{Sui et~al.(2018)Sui, Yu \& Luo}]{Sui2018}
\bibinfo{author}{Sui, Y.}, \bibinfo{author}{Yu, W.}, \& \bibinfo{author}{Luo,
  Q.} (\bibinfo{year}{2018}).
\newblock \bibinfo{title}{{Jointly Optimized Extreme Learning Machine for
  Short-Term Prediction of Fading Channel}}.
\newblock {\it \bibinfo{journal}{IEEE Access}\/},  {\it \bibinfo{volume}{6}\/},
  \bibinfo{pages}{49029--49039}. \DOIprefix\doi{10.1109/ACCESS.2018.2868480}.
\bibitem[{Tan \& Hirose(2009)}]{Tan2009}
\bibinfo{author}{Tan, S.}, \& \bibinfo{author}{Hirose, A.}
  (\bibinfo{year}{2009}).
\newblock \bibinfo{title}{{Low-calculation-cost fading channel prediction using
  chirp Z-transform}}.
\newblock {\it \bibinfo{journal}{Electronics Letters}\/},  {\it
  \bibinfo{volume}{45}\/}, \bibinfo{pages}{418}.
  \DOIprefix\doi{10.1049/el.2009.3472}.
\bibitem[{Tibshirani(1996)}]{Tibshirani1996}
\bibinfo{author}{Tibshirani, R.} (\bibinfo{year}{1996}).
\newblock \bibinfo{title}{{Regression Shrinkage and Selection Via the Lasso}}.
\newblock {\it \bibinfo{journal}{Journal of the Royal Statistical Society:
  Series B (Methodological)}\/},  {\it \bibinfo{volume}{58}\/},
  \bibinfo{pages}{267--288}.
  \DOIprefix\doi{10.1111/j.2517-6161.1996.tb02080.x}.
\bibitem[{Trabelsi et~al.(2018)Trabelsi, Bilaniuk, Zhang, Serdyuk, Subramanian,
  Santos, Mehri, Rostamzadeh, Bengio \& Pal}]{Trabelsi2018}
\bibinfo{author}{Trabelsi, C.}, \bibinfo{author}{Bilaniuk, O.},
  \bibinfo{author}{Zhang, Y.}, \bibinfo{author}{Serdyuk, D.},
  \bibinfo{author}{Subramanian, S.}, \bibinfo{author}{Santos, J.~F.},
  \bibinfo{author}{Mehri, S.}, \bibinfo{author}{Rostamzadeh, N.},
  \bibinfo{author}{Bengio, Y.}, \& \bibinfo{author}{Pal, C.~J.}
  (\bibinfo{year}{2018}).
\newblock \bibinfo{title}{{Deep Complex Networks}}.
\newblock In {\it \bibinfo{booktitle}{Proceedings of ICLR 2018}\/}
  \bibinfo{number}{2016}
\newblock (pp. \bibinfo{pages}{1--19}).
  \href{http://arxiv.org/abs/1705.09792}{\tt arXiv:1705.09792}.
\bibitem[{{Tzyy-Chyang Lu} et~al.(2013){Tzyy-Chyang Lu}, {Gwo-Ruey Yu} \&
  {Jyh-Ching Juang}}]{Tzyy-ChyangLu2013}
\bibinfo{author}{{Tzyy-Chyang Lu}}, \bibinfo{author}{{Gwo-Ruey Yu}}, \&
  \bibinfo{author}{{Jyh-Ching Juang}} (\bibinfo{year}{2013}).
\newblock \bibinfo{title}{{Quantum-Based Algorithm for Optimizing Artificial
  Neural Networks}}.
\newblock {\it \bibinfo{journal}{IEEE Transactions on Neural Networks and
  Learning Systems}\/},  {\it \bibinfo{volume}{24}\/},
  \bibinfo{pages}{1266--1278}. \DOIprefix\doi{10.1109/TNNLS.2013.2249089}.
\bibitem[{Valle(2014)}]{Valle2014}
\bibinfo{author}{Valle, M.~E.} (\bibinfo{year}{2014}).
\newblock \bibinfo{title}{{Complex-Valued Recurrent Correlation Neural
  Networks}}.
\newblock {\it \bibinfo{journal}{IEEE Transactions on Neural Networks and
  Learning Systems}\/},  {\it \bibinfo{volume}{25}\/},
  \bibinfo{pages}{1600--1612}. \DOIprefix\doi{10.1109/TNNLS.2014.2341013}.
\bibitem[{Wang et~al.(2018)Wang, Xu, Yang \& Zurada}]{Wang2018}
\bibinfo{author}{Wang, J.}, \bibinfo{author}{Xu, C.}, \bibinfo{author}{Yang,
  X.}, \& \bibinfo{author}{Zurada, J.~M.} (\bibinfo{year}{2018}).
\newblock \bibinfo{title}{{A Novel Pruning Algorithm for Smoothing Feedforward
  Neural Networks Based on Group Lasso Method}}.
\newblock {\it \bibinfo{journal}{IEEE Transactions on Neural Networks and
  Learning Systems}\/},  {\it \bibinfo{volume}{29}\/},
  \bibinfo{pages}{2012--2024}. \DOIprefix\doi{10.1109/TNNLS.2017.2748585}.
\bibitem[{Yuan \& Lin(2006)}]{Yuan2006}
\bibinfo{author}{Yuan, M.}, \& \bibinfo{author}{Lin, Y.}
  (\bibinfo{year}{2006}).
\newblock \bibinfo{title}{{Model selection and estimation in regression with
  grouped variables}}.
\newblock {\it \bibinfo{journal}{Journal of the Royal Statistical Society:
  Series B (Statistical Methodology)}\/},  {\it \bibinfo{volume}{68}\/},
  \bibinfo{pages}{49--67}. \DOIprefix\doi{10.1111/j.1467-9868.2005.00532.x}.
\bibitem[{Zhao et~al.(2017)Zhao, Gao, Beaulieu, Chen \& Ji}]{Zhao2017}
\bibinfo{author}{Zhao, Y.}, \bibinfo{author}{Gao, H.},
  \bibinfo{author}{Beaulieu, N.~C.}, \bibinfo{author}{Chen, Z.}, \&
  \bibinfo{author}{Ji, H.} (\bibinfo{year}{2017}).
\newblock \bibinfo{title}{{Echo State Network for Fast Channel Prediction in
  Ricean Fading Scenarios}}.
\newblock {\it \bibinfo{journal}{IEEE Communications Letters}\/},  {\it
  \bibinfo{volume}{21}\/}, \bibinfo{pages}{672--675}.
  \DOIprefix\doi{10.1109/LCOMM.2016.2632120}.
\bibitem[{Zou \& Hastie(2005)}]{Zou2005}
\bibinfo{author}{Zou, H.}, \& \bibinfo{author}{Hastie, T.}
  (\bibinfo{year}{2005}).
\newblock \bibinfo{title}{{Regularization and variable selection via the
  elastic net}}.
\newblock {\it \bibinfo{journal}{Journal of the Royal Statistical Society:
  Series B (Statistical Methodology)}\/},  {\it \bibinfo{volume}{67}\/},
  \bibinfo{pages}{301--320}. \DOIprefix\doi{10.1111/j.1467-9868.2005.00503.x}.

\end{thebibliography}

\end{document}